\documentclass[aps,prl,twocolumn,showpacs,superscriptaddress,floatfix,nofootinbib]{revtex4-1}
\usepackage{graphicx,graphics}
\usepackage{dcolumn}
\usepackage{amsmath,amssymb,amsfonts}
\usepackage{latexsym,verbatim}
\usepackage{bm}
\usepackage{color}
\usepackage[breaklinks=true,colorlinks,citecolor=blue,linkcolor=blue,urlcolor=blue]{hyperref}
\usepackage{tabularx}

\newcommand{\be}{\begin{equation}}
\newcommand{\ee}{\end{equation}}
\newcommand{\bea}{\begin{eqnarray}}
\newcommand{\eea}{\end{eqnarray}}
\newcommand{\nn}{\nonumber}

\begin{document}
\title{Deterministic chaos and fractal entropy scaling in Floquet CFT}

\author{Dmitry S. Ageev}
\email{ageev@mi-ras.ru}
\affiliation{Department of Mathematical Methods for Quantum Technologies, Steklov Mathematical Institute, Russian Academy of Sciences, Gubkin str. 8, 119991
Moscow, Russia}
\author{Andrey A. Bagrov}
\email{andrey.bagrov@physics.uu.se}
\affiliation{Department of  Physics  and  Astronomy,  Uppsala  University, Box 516,  SE-75120  Uppsala,  Sweden}
\affiliation{Theoretical Physics and Applied Mathematics Department, Ural Federal University, Mira Str. 19, 620002 Ekaterinburg, Russia}

\author{Askar A. Iliasov}
\email{a.iliasov@science.ru.nl}
\affiliation{Institute for Molecules and Materials, Radboud University, Heyendaalseweg 135, 6525AJ Nijmegen, The Netherlands}
\affiliation{Space Research Institute of the Russian Academy of Science, 117997 Moscow, Russia}

\begin{abstract}

In this paper, we study 2d Floquet conformal field theory, where the external periodic driving is described by iterated logistic or tent maps. These maps are known to be typical examples of dynamical systems exhibiting the order-chaos transition, and we show that, as a result of such driving, the entanglement entropy scaling develops fractal features when the corresponding dynamical system approaches the chaotic regime. For the driving set by the logistic map, fractal contribution to the scaling dominates, making entanglement entropy highly oscillating function of the subsystem size.
\end{abstract}

\maketitle
The concept of Floquet driving keeps attracting considerable interest in different areas of quantum many-body theory since it  provides a powerful approach to controllable manipulation of phases of matter. In the study of topologically protected states of matter \cite{Jiang:2011xv, Wieder:2020lvf}, many-body localization \cite{ab1,ab2}, and thermalization \cite{Maffei:2018knw, rig1}, it was shown to be a tool for both externally tuning conventional phase transitions \cite{Stepanov} and engineering novel phases that are non-existent in equilibrium \cite{Po:2016qlt}.

A natural playground to address the dynamics of periodically modulated correlated systems is the framework of Floquet conformal field theories. Recently, a wide class of Floquet modulated 2d CFTs has been proven to be exactly solvable \cite{Wen:2018agb,Fan:2019upv,Lapierre:2019rwj}. In the proposed setting, one drives the system by periodically interchanging two Hamiltonians, namely, the canonical Hamiltonian of 2d CFT on a strip, and its deformation. While originally the particular case of the so-called sine-square deformation (SSD) has been considered \cite{Wen:2018agb, Hikihara:2011}, later on, it was realized that dynamics of inhomogenous CFT \cite{MacCormack:2018rwq,Wen:2016inm,Langmann:2019nna,Caputa:2020mgb} is exactly solvable for general deformations \cite{Moosavi:2019fas,Gawedzki:2017woc}.
Since already in the simplest SSD case, the driving leads to non-trivial energy flows in the system causing localization of energy density, which pattern resembles black-hole-like structures \cite{Fan:2019upv,Lapierre:2019rwj,coh}, it is tempting to exploit the approach in its full generality and investigate what types of unconventional emergent phenomena can occur in the Floquet driven systems.

 \begin{figure*}[t!]
        \centering
        \includegraphics[height=5cm]{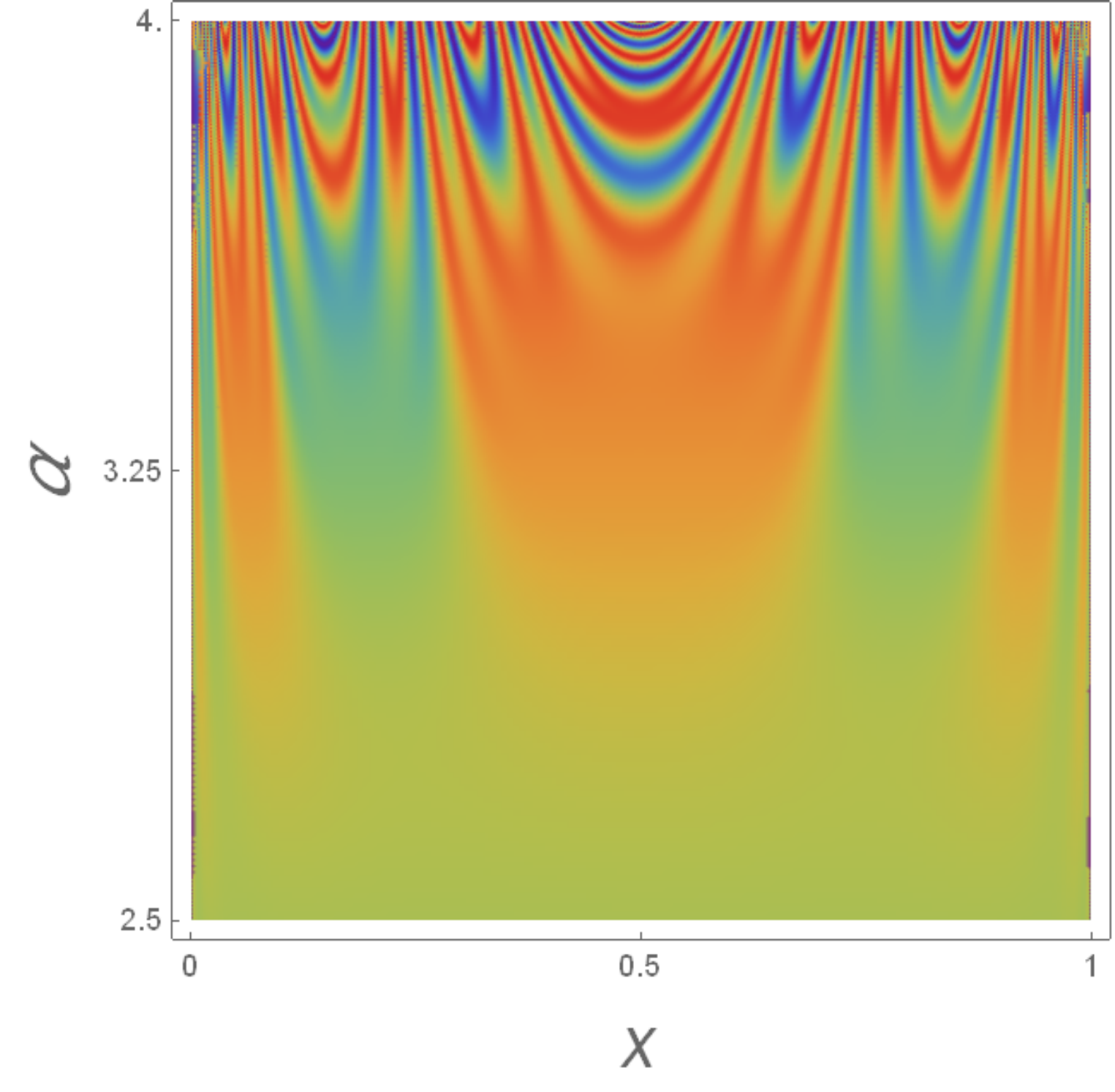}
        \includegraphics[height=5cm]{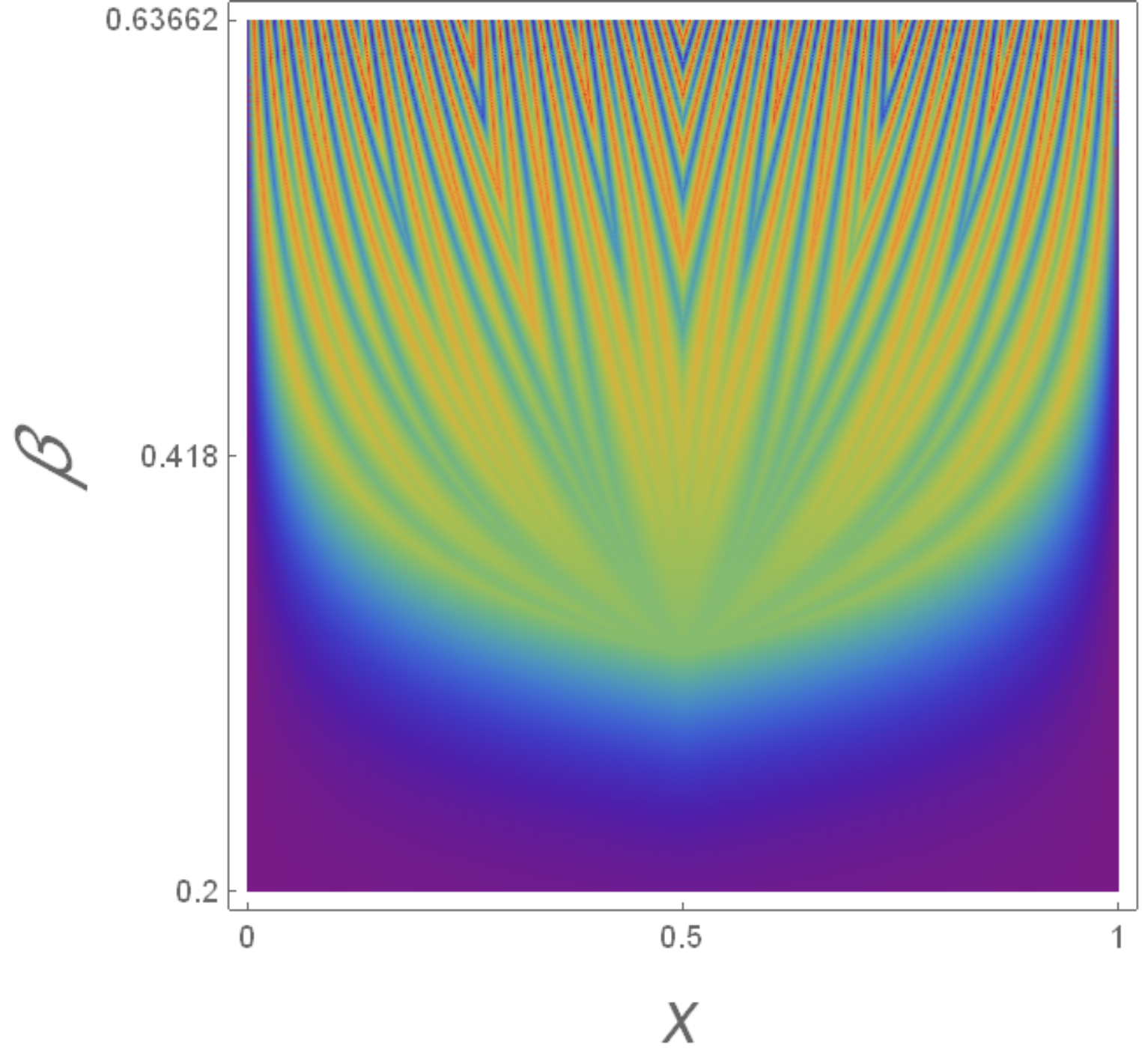}

        \caption{\label{fig:dynsys} The dependence of iterated maps $z_n$ (taken on the real axis; $n=10$) on parameters $\alpha$ and $\beta$ for the  logistic map (left) and the tent map (right).}
    \end{figure*}

To proceed along this line, one should bear in mind that the Floquet theory as a branch of mathematics can be viewed as a part of a more general theory of classical dynamical systems \cite{Brown:DynSys,dyn1,dyn2}.
In the previous works \cite{Wen:2018agb,Fan:2019upv}, the following correspondence
  between driven two-dimensional conformal field theories and dynamical systems has been established implicitly:
 \begin{gather}
\text{Evolution of a deformed 2d CFT} \nn \\ \big\Updownarrow \nn \\ \text{Holomorphic dynamics of iterated maps} \nn
 \end{gather}
Dynamical systems defined by iterative composition of simple maps are known to exhibit chaos, bifurcations and self-organization phenomena.
For many standard choices of the holomorphic maps, evolution of the corresponding dynamical system generates fractals.
 Fractals, which are objects of non-integer dimension, have intriguing non-smooth behavior and can appear even in the dynamics of seemingly regular structures.
 A canonical example of a fractal is the Julia set that can emerge as the spectrum of quasiperiodic (the famous Hofstadter butterfly) or self-similar potentials \cite{Hofstadter:1976, Kohmoto:1983, Domany_1983}. In fact, the emergence of fractals \cite{ratiter} is a commonplace for a general choice of iterated functions. Many studies are devoted to understanding fractal structures in cold-atom systems, topological insulators, mesoscopic systems etc. \cite{Shang_etal:2015, Kempkes_etal:2018, Gubser:2018, Westerhout:2018, Pal_etal:2019, IlKatSheng:2019, IlKatSheng:2020, IlKatSheng:2020b}.

Treating the Floquet driving of a CFT as a dynamical system, which can manifest chaotic or fractal properties on the level of map iterations, does not evidently correspond to the common approach normally taken in quantum chaos theory \cite{chaosrmt1,chaosrmt2}. Hence it is interesting to elucidate the connection between chaotic dynamics in quantum systems and the more classical theory of dynamical systems.
 In this paper, we report that periodic Floquet modulation based on a certain choice of iterated holomorphic maps can evoke fractal patterns of quantum correlations in the driven state. Namely, it leads to emergence of an exotic class of quantum states with {\it fractal scaling law} of entanglement entropy.
To achieve that, we consider CFT deformations that can be described in terms of dynamical systems generated by either logistic or tent map. Both these maps are extremely simple: the tent map is a discontinuous linear function and the logistic map is a quadratic function. Still, they lead to highly non-trivial dynamics exhibiting transition to the regime of deterministic chaos, -- irregular dynamics governed by deterministic laws that are highly sensitive to the initial conditions \cite{dyn3}.

Following the proposal of \cite{Wen:2018agb,Wen:2018vux}, we start with a theory defined on the strip of width $L$ in the complex plane:
\be \label{H0}
H=\int_0^L \frac{d\sigma}{2\pi}\Big(f(w)T(w)+f(\bar w) \bar T(\bar w)   \Big),
\ee
where $w=\tau+ ix$, $x \in (0,L)$\footnote{After analytical continuation, $x$ would be the spatial coordinate.}. Here $T(w)$ and $\bar T(\bar w)$ are the holomorphic and anti-holomorphic components of the stress-energy tensor correspondingly, and $f(w)$ defines the deformation, with $f(w)=1$ corresponding to the undeformed CFT.

To introduce the Floquet driving, we first perform a conformal map from the strip onto the complex plane with a slit by
\be
z=\exp\left(\frac{2\pi}{L}w\right),
\ee
and then make change of variables $\chi=\chi(z)$
\be \label{eq:chi-f}
\chi(z)=\exp\left(\int\frac{dz}{zf(z)}\right),
\ee
bringing \eqref{H0}  to the uniform Hamiltonian \cite{Fan:2019upv}:
\be
H=\frac{2\pi}{L}\int_{\cal C}  \frac{\chi}{2\pi i}T(\chi)d\chi+ c.c.+...
\ee
where a constant term resulting from the transformation is omitted.
In these coordinates, the  Hamiltonian evolution in Euclidean time $\tau$ is given simply by the dilatation transformation $\chi\rightarrow \lambda\cdot \chi$ with $\lambda=e^{\frac{2 \pi  }{L}\tau}$ (through the rest of the paper, we take $L=1$). In $z$ coordinates, this implies that the
evolution in Euclidean time for some $\tau$ governed by the deformed Hamiltonian \eqref{H0} reduces to a change of variables $z_1(z)$ obeying the identity
\be\label{eq:chn}
\chi(z_{1})=\lambda \chi(z),
\ee
which is the Schr\"oder spectral functional equation from the theory of dynamical systems \cite{SM}.
The Hamiltonian evolution for time $n\cdot \tau$ is then given by composition
\be\label{eq:zn}
z_n(z)=\underbrace{(z_1\circ z_1 \ldots \circ z_1)}_{n\,\, times}(z).
\ee
From this, it readily follows that the Floquet driving of a CFT, when the system Hamiltonian is periodically swung with time step $\tau$, can be described by alternating composition of two different
functions $z_1(z)$ and $\widetilde z_1(z)$:
\be\label{eq:zn2}
z_n(z)=\underbrace{(z_1\circ \widetilde z_1 \ldots \circ z_1)}_{n\,\, times}(z),
\ee
where each of the functions corresponds to its own deformation $f(z)$.

We will be working with two dynamical systems defined by the logistic map
\bea \label{logistic}
&&z_1^{\text{log}}(x)=\alpha x (1-x)
\eea
and the tent map
\be \label{tent}
z_1^{\text{tent}}(x)=\begin{cases}
\beta x, \, x<\frac{1}{2}\\
\beta (1-x), \, x>\frac{1}{2},
\end{cases}
\ee
which can also be equivalently represented as
\be\label{zntent}
z^{tent}_1(x)=\frac{\beta}{\pi}\arcsin(\sin(\pi x)),
\ee
where $x\in(0,1)$ \footnote{Eq. \eqref{zntent} allows us to continue the tent map to the complex plane: $f(z)=\arcsin(z)$ is expressed through complex logarithms.}.
The Floquet evolution of the CFT is then given by iterated composition \eqref{eq:zn} of maps \eqref{logistic} or \eqref{tent} and the dilatation.
For both choices, it can be shown that by properly rescaling the holomorphic map, the stroboscopic dynamics can in fact be reformulated as a Hamiltonian dynamics following single quench \cite{SM}.

\begin{figure*}[t!]
        \centering

        \includegraphics[width=5.7cm]{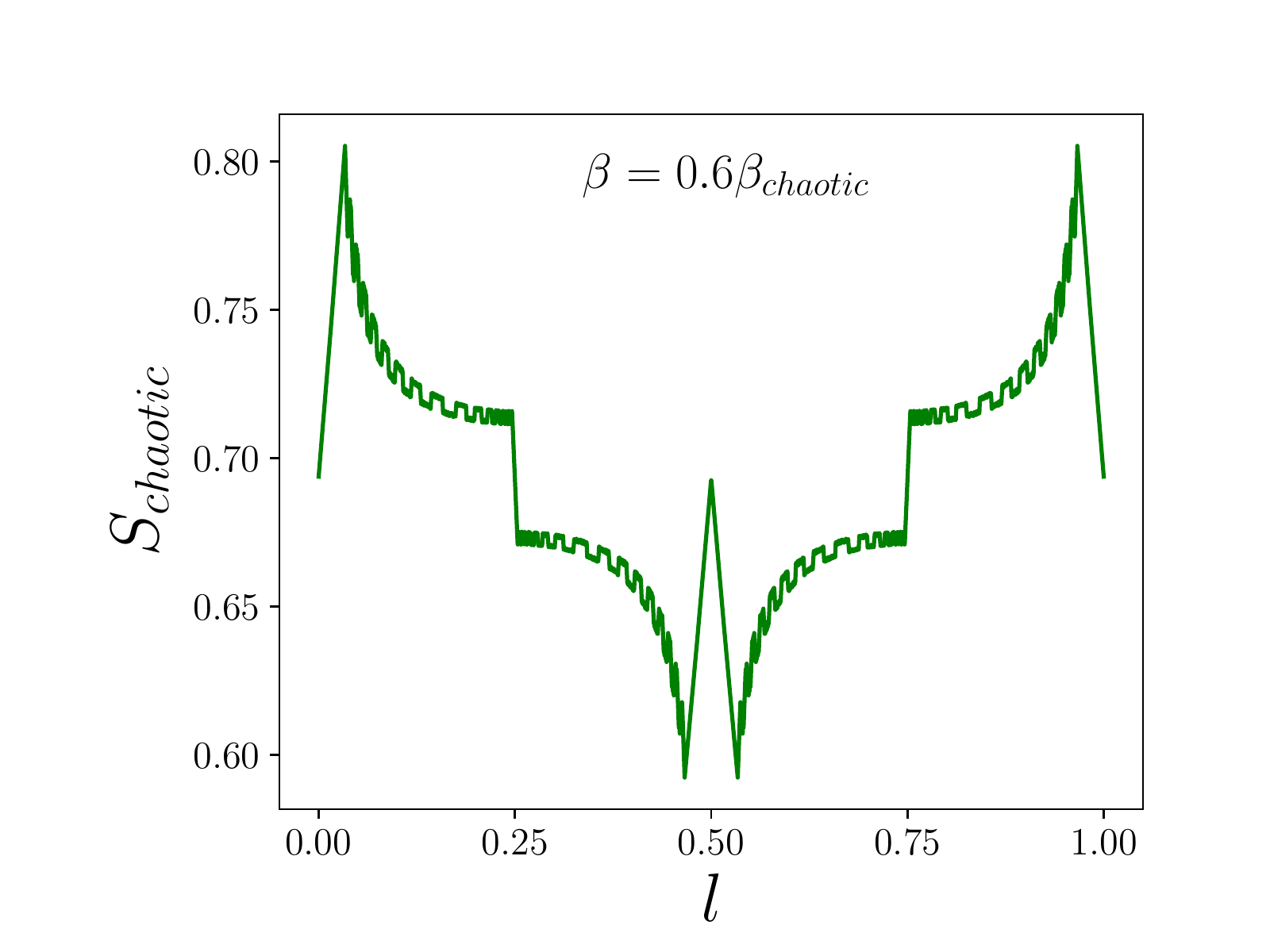}
        \includegraphics[width=5.7cm]{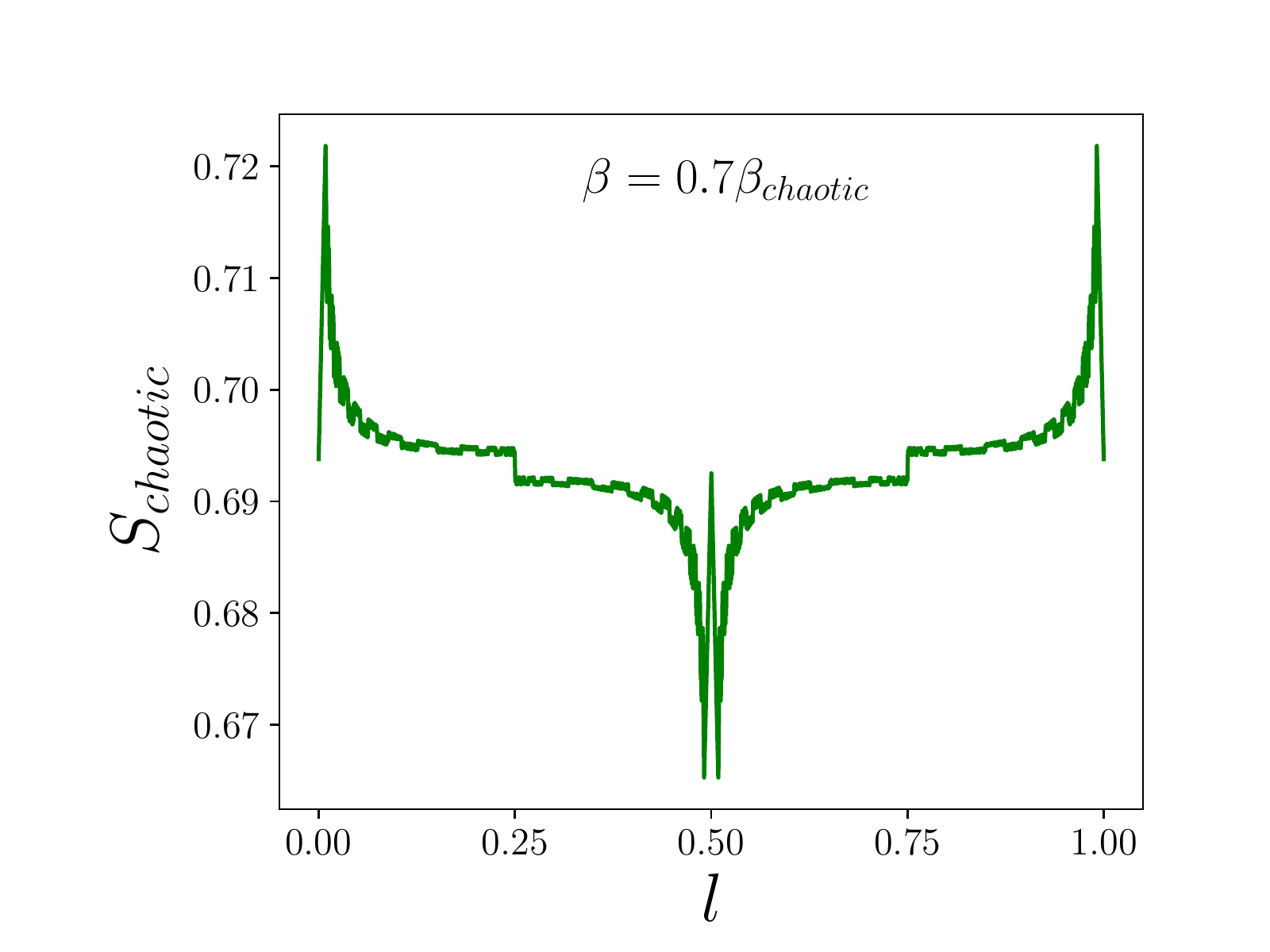}
        \includegraphics[width=5.7cm]{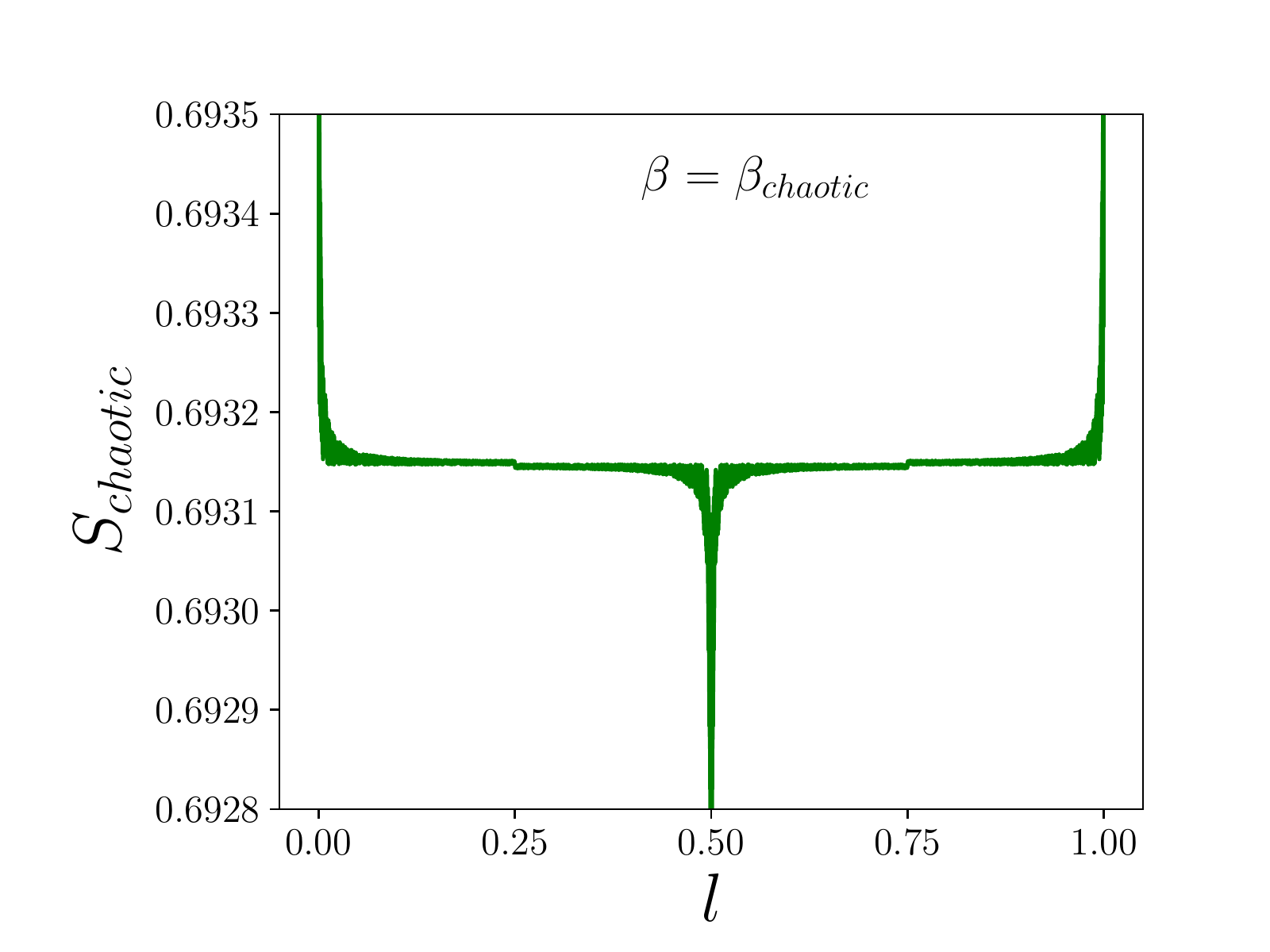} \\

        \includegraphics[width=5.7cm]{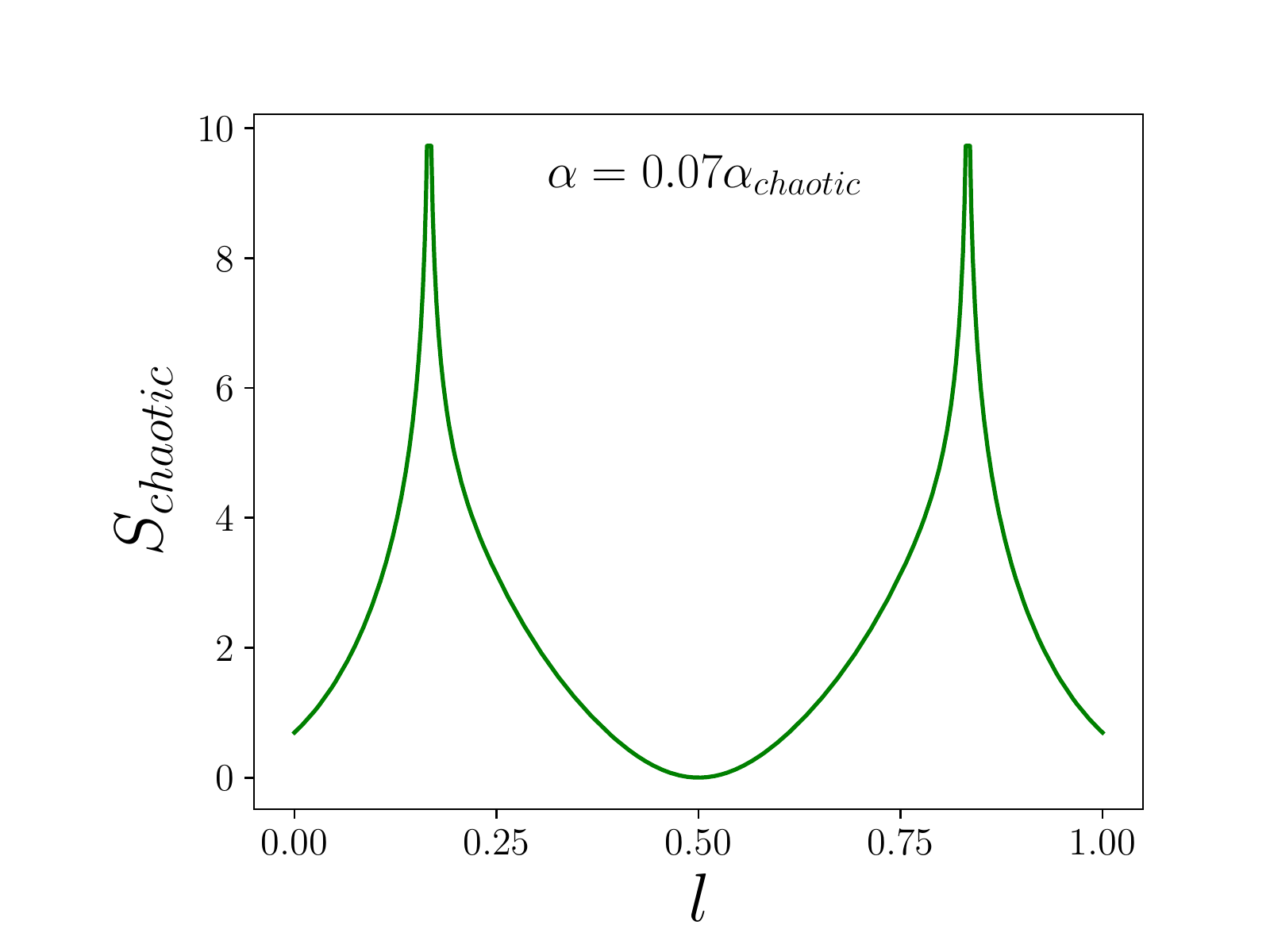}
        \includegraphics[width=5.7cm]{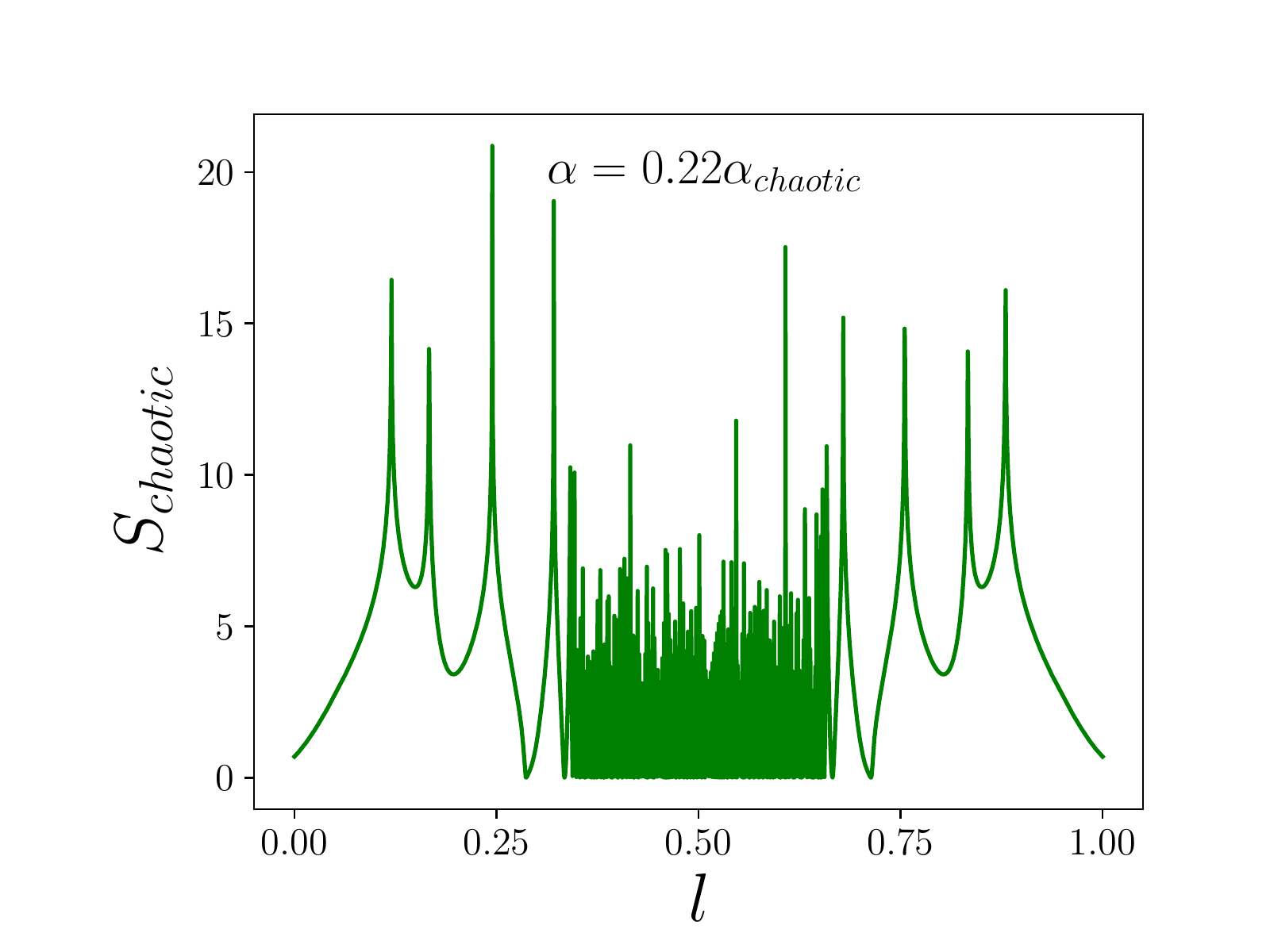}
        \includegraphics[width=5.7cm]{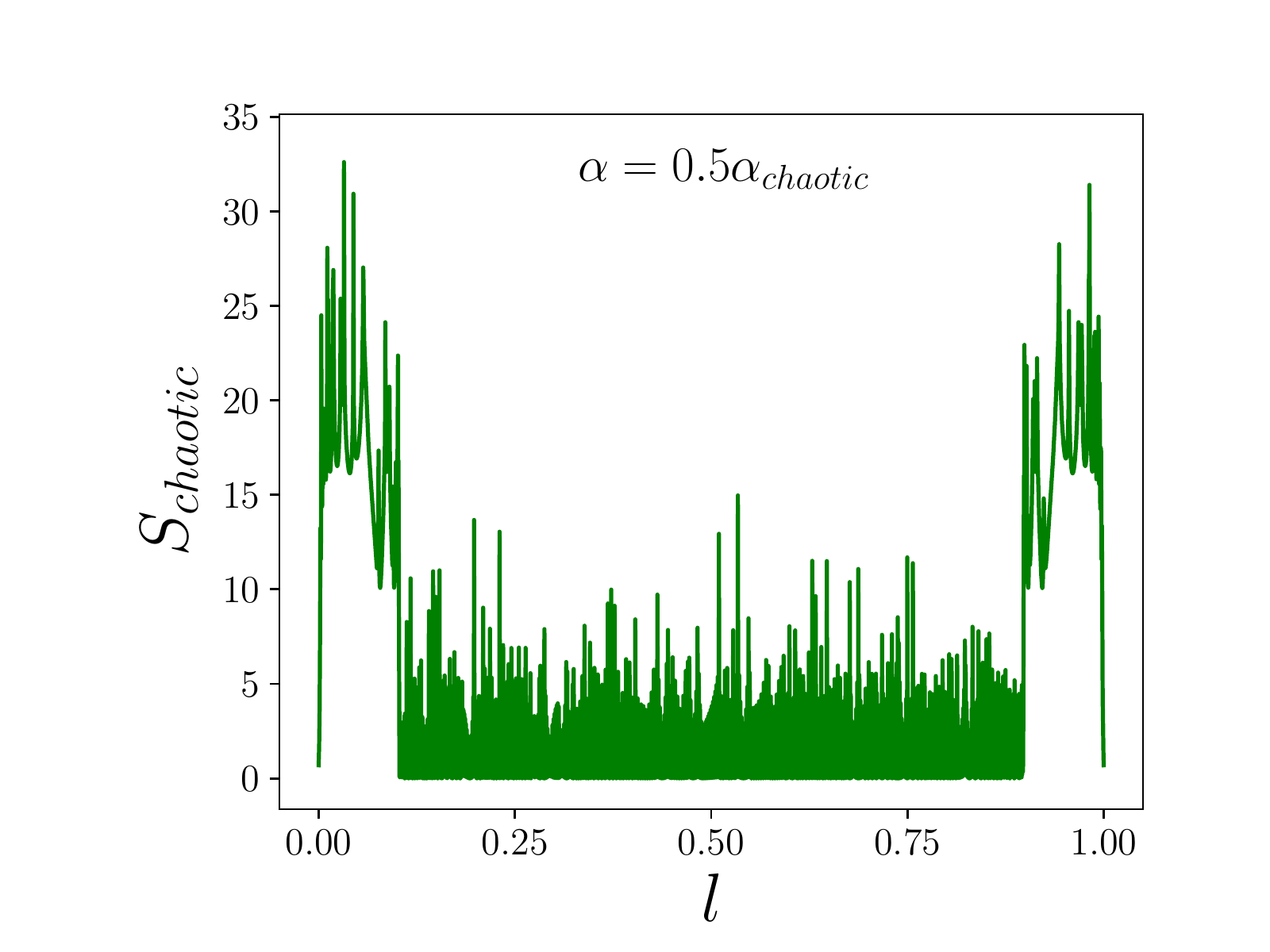}

        \caption{\label{fig:tentSnorm-alpha} Irregular part of the entanglement entropy $S_{chaotic}$ as a function of bipartition coordinate. Note that for the purpose of nicer representation we define $\Delta S = -(S_{reg}+S_{chaotic})$, so that in fact the contributions to the entanglement entropy are negative. Top row:  after $n=18$ Floquet steps of the tent map. Bottom row: after $n=10$ steps of the logistic map.}
    \end{figure*}

Iterative dynamics of these maps gradually becomes more and more complicated when $\alpha$ and $\beta$ parameters approach the critical values $\alpha_{chaotic}=4$ and $\beta_{chaotic}=2$ corresponding to the truly chaotic regime. One can naturally expect that properties of a CFT driven by such a dynamical system would encode the deterministic chaos in a non-trivial way.
To study that, for the Floquet driving Eq. \eqref{eq:zn2}, we choose $z_1$ to be either the tent or logistic map, and $\widetilde z_1$ -- the dilatation transformation.

As shown in \cite{SM}, in order to describe evolution of a CFT with some deformation $f$, one-step transformation $z_1$ should allow Schr{\"o}der equation \eqref{eq:chn}
to have non-trivial solutions. $f(z)$ can then be restored from \eqref{eq:chi-f}.
In general, it is hard to find the solution for $\chi$ analytically, and the answer is known only for a limited number of functions $z_1$ \cite{Curtright:2009,Milnor:2004}. In the chaotic regime, for the logistic and tent maps,  $\chi(z)$ is
\cite{CollJeanPierre:1980}
\bea
&&\text{logistic, }\alpha=4:\,\,\chi(z)=\arcsin^2(\sqrt{z}),\,\,\,\,\,\,\\
&&\text{tent, }\beta=2:\,\,\chi(z)=\arcsin^2\left(\sin \left(\frac{\pi  z}{2}\right)\right), \nn
\eea
with the resulting expressions for Hamiltonian deformations:
\bea
&&f_{\alpha=4}(z)=\sqrt{\frac{1}{z}-1}\arcsin(\sqrt{z}),\,\,\,\,\,\,\\
&&f_{\beta=2}(z)=\frac{1}{\pi z}\arcsin\left(\sin \left(\frac{\pi  z}{2}\right)\right)\cdot \frac{\sqrt{\cos^2 \left(\frac{\pi  z}{2}\right)}}{\cos \left(\frac{\pi  z}{2}\right)}. \nn
\eea
For both maps, the corresponding eigenvalue is $\lambda=4$.

For the logistic map, it turns out to be possible to obtain an explicit form of the Hamiltonian deformation also for $\alpha=2$:
\begin{equation}
    f(z) = \left(1-\frac{1}{2z} \right) \log(1-2z),
\end{equation}
as well as an approximate expression at small values of $\alpha$:
\begin{equation}
    f(z) = \frac{(1-z)(1-\alpha z(1-z))}{(1-2z)(1-2\alpha z(1-z))}.
\end{equation}
We provide idea of the derivation in \cite{SM}. For the tent map, to reconstruct deformations of the Hamiltonian, one can utilize the expressions corresponding to logistic map and employ the Milnor-Thurston kneading theory that connects logistic and tent maps by a semiconjugate relation \cite{MilnThurst:1988} (we outline basic elements of this construction in \cite{SM}). Fortunately, within the approach we have taken here, it is possible to proceed without having explicit expressions for $f(z)$.

As we show in \cite{SM}, CFT with deformations of this type are non-Hermitian. Recently, theories of this kind have attracted interest because of their relevance in the context of non-unitary dynamics of open quantum systems \cite{Saleur, Moosavi}. Experimentally, periodic Floquet driving that induces non-Hermitian evolution can be implemented by switching on and off the coupling between the quantum system and an external bath \cite{WuAn, GongWang}.

To derive real-time dynamics of the CFT, it is natural to get back to the original $w$-coordinates on the strip.
The theory is defined on spatial interval $x\in (0,1)$, so we divide it into two subregions $(0,\ell)$ and $(\ell,1)$, and compute evolution of the von Neumann entanglement entropy of this bipartition, which can be expressed as the limit \be
S(\ell,t)=\underset{\nu\rightarrow 1}{\text{lim}} S^{\nu}(\ell,t),\,\,\,\, S^{\nu}(\ell,t)=\frac{1}{1-\nu}\log \text{tr}\left[\rho^\nu(\ell,t)\right],
\ee
where $S^{\nu}(\ell,t)$ is the Renyi entropy.
The trace $\text{tr}\left[\rho^\nu(\ell,t)\right]$ can be computed as one-point correlation function of primary twist operator $\mathcal{T}_{\nu}$ of conformal dimension $h$ \cite{Calabrese:2004eu,Calabrese:2009qy}: \be
\text{tr}\left[\rho^\nu(\ell,t)\right]=\langle\psi(t)|\mathcal{T}_{\nu}(\ell)|\psi(t) \rangle,\,\,\,\,h=\frac{c}{24}\left(\nu-\frac{1}{\nu}\right).
\ee
Here $|\psi(t)\rangle$ is the state of the system after $n$ cycles of driving ($t=n\tau$).
The explicit expression for this correlation function is \cite{Calabrese:2009qy}
\bea
\langle\psi(t)|\mathcal{T}_{\nu}(\ell)|\psi(t) \rangle=A_{\nu}\left(\frac{\partial z_n}{\partial w}\right)^{h}\left(\frac{\partial\bar z_n}{\partial \bar w}\right)^{h} \cdot \\ \cdot \nn \left(\frac{1}{4\sqrt{z_n \bar z_n}}\right)^{h}\left(\frac{2 i \epsilon }{\sqrt{z_n}-\sqrt{\bar z_n}}\right)^{2h},
\eea
where $\varepsilon$ is a UV cutoff, and $A_{\nu}$ is a constant that drops out of the subsequent calculations.
After taking the $\nu~\rightarrow~1$ limit and subtracting the ground state entanglement and the constant terms, we obtain
\be
\Delta S(\ell,t)=-\frac{ c}{12}
\log\left(-\frac{\partial_w z_n \partial_{\bar w} \bar z_n}{\sqrt{z_n \bar z_n }(\sqrt{z_n}-\sqrt{\bar z_n})^2}\right).
\ee
To extract the fractal contribution to the entropy scaling, it is convenient to use ``radial'' parametrization ${z_n=R_n\cdot \exp( i\phi_n)}$ and ${\bar z_n=R_n\cdot \exp( -i\phi_n)}$ \cite{Wen:2018agb}. In these variables, we get
\begin{gather} \label{SN}
\Delta S=-\left(S_{reg}+S_{chaotic}\right),\,\,\,\, \\ S_{reg}=\log\frac{\partial_w z_n\partial_{\bar w}\bar z_n  }{4 R_n^2}, \,\,\,\, S_{chaotic}=\log{\csc ^2\left(\frac{\phi_n}{2}\right)}, \nn
\end{gather}
where we imposed $c=12$ for the sake of nice normalization. The names ``regular'' and ``chaotic'' come from the fact that for both studied types of Floquet driving, $R_n$ and the determinant $\partial_w z_n\partial_{\bar w}\bar z_n$  are regular objects, while $\csc ^2\left(\phi_n/2\right)$ that depends on phase $\phi_n$ exhibits erratic fractal/chaotic behaviour.

\begin{figure}[t!]
        \centering

        \includegraphics[width=7.5cm]{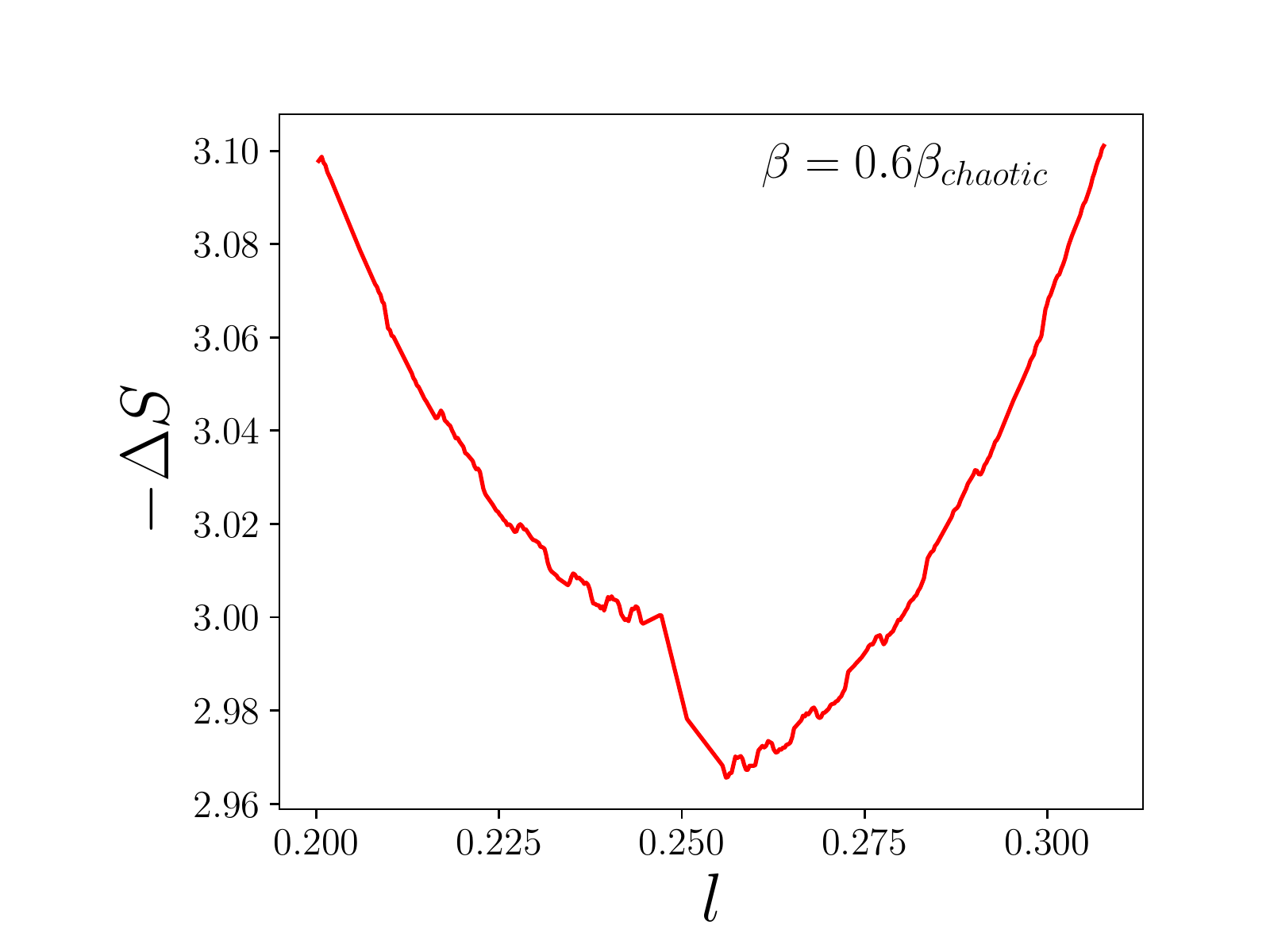}
        \includegraphics[width=7.5cm]{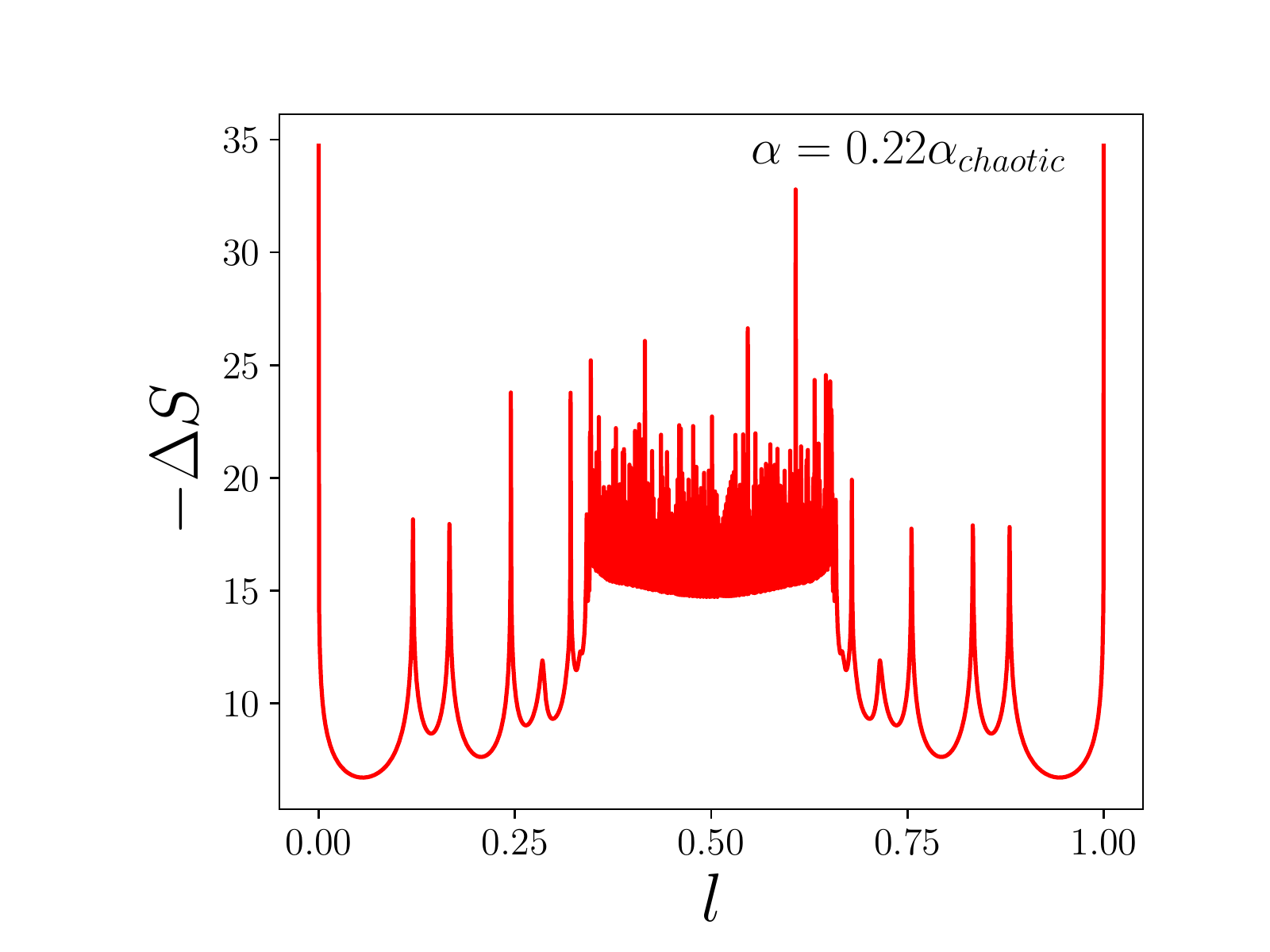}

        \caption{\label{fig:fullS} Full correction to the entanglement entropy $-\Delta S=S_{reg}+S_{chaotic}$ (taken with overall negative sign) as a function of bipartition coordinate for the tent (top, after $n=18$ driving steps) and the logistic maps (bottom, after $n=10$ driving steps).}
    \end{figure}

For the tent map driving,  the fractal pattern in $S(\ell,t)$ is explicit, but its contribution to the overall entropy is rather small, especially upon approaching the chaotic regime, top panels of Figs. \ref{fig:tentSnorm-alpha} and \ref{fig:fullS}. Away from $\beta_{chaotic}$, it has a shape of fractal with rather low Hausdorff dimension.
For the logistic map driving, the irregular contribution acquires much higher amplitude and becomes comparable with the regular part, which allows one to speak about {\it fractal scaling} of the entanglement entropy\footnote{States of a closed CFT obey strong subadditivity condition, which manifests itself on the level of entanglement entropy scaling as concavity of $S(l)$: $S''(l)\,\, \leq \,\, 0\,\, \forall\,\, l$ \cite{HeadrickTakayanagi}. Given that our setting is non-Hermitian and can possibly be interpreted as an open quantum system, this constraint can be naturally violated.}, bottom panels of Figs. \ref{fig:tentSnorm-alpha} and \ref{fig:fullS}.

\begin{figure}[t!]
        \centering

        \includegraphics[width=7.5cm]{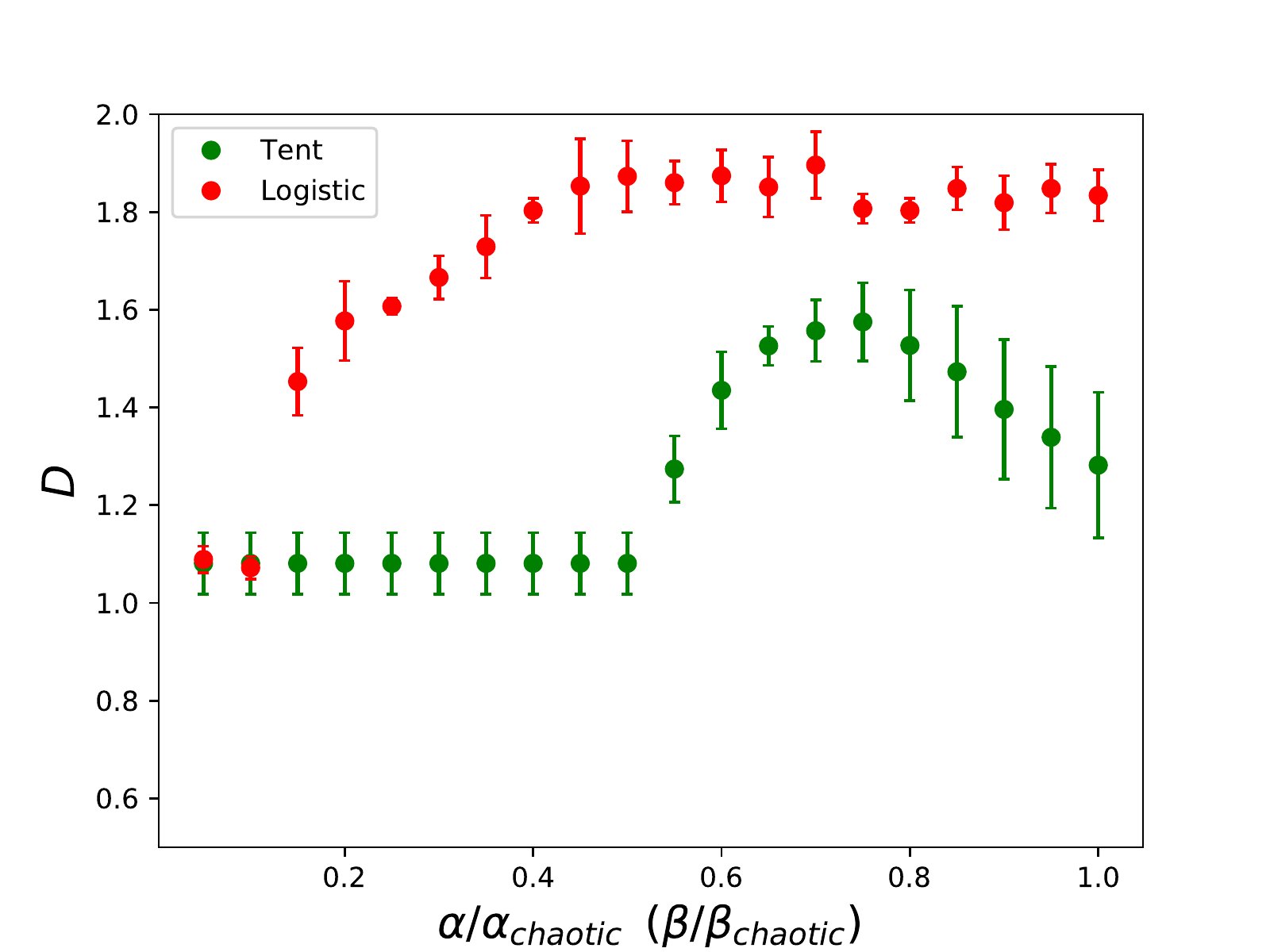}

        \caption{\label{fig:Hausdorff} Dependence of Hausdorff dimension of $S_{chaotic}$ on the value of driving parameter $\alpha/\alpha_{chaotic}$ ($\beta/\beta_{chaotic}$) for the tent and logistic drivings. For every value of $\alpha$ ($\beta$), the dimension and the error bars are correspondingly mean and standard deviation of so-called local fractal dimensions computed for different windows of spatial scales ranging from $\Delta l \in (0.0004,0.002)$ to $\Delta l \in (0.038,0.04)$ \cite{Dubuc, SM}.}
    \end{figure}

While self-similarity of the entropy curves is visually recognizable, to really prove that $S_{chaotic}$ has a fractal shape, it is instructive to compute its Hausdorff dimension. Since here we are dealing with one-dimensional curves which cannot be viewed as sets embedded in a two-dimensional space with both coordinates having the meaning of length, naively applying box-counting algorithm would lead to meaningless results. Instead, we use the method suggested in \cite{Dubuc} (see also \cite{Pilgrim}) designed specifically for computing fractal dimensions of one-dimensional profiles. We describe it in detail in \cite{SM}, and here provide the results. In Fig.\ref{fig:Hausdorff}, we plot the dependence of fractal dimension of $S_{chaotic}$. One can see that, in the logistic driving case, significant fractality develops already for small values of $\alpha$, while for the tent map it is not until $\beta=0.5\beta_{chaotic}=1$ when the entropy profile acquires fractal features. Error bars reflect how well the estimated value persists across a range of spatial scales. In other words, small error bars mean that the corresponding profile is truly self-similar, while large error bars, like for $\beta/\beta_{chaotic}>0.75$, signalize that the fractal dimension in fact flows.
In both cases, upon $\alpha\rightarrow\alpha_{\text{chaotic}}$ and $\beta\rightarrow\beta_{\text{chaotic}}$, the irregular part of entanglement entropy gradually approaches a fully chaotic regime passing through fractal configurations at smaller values of the parameters. It must be stressed out that, for a given choice of $\alpha$ or $\beta$, the entropy profile tends to stabilize as $n\rightarrow \infty$ since corresponding iterative conformal map \eqref{eq:zn2} stabilizes in this limit. This observation is important in two regards. First, for large enough times one can view the evolving stroboscopic state as effectively stationary, making this kind of Floquet driving a candidate mechanism for creating {\it stable in time} fractal phases of matter that are not achievable in static systems. Secondly, the sharp and distinguishable fractal structures at $\alpha \ll \alpha_{chaotic} \,\,\beta \ll \beta_{chaotic}$, that could be of higher potential interest than the pseudo-random profiles emerging in the chaotic phase, do not gain ``featureless'' corrections over the course of the repeated driving.
\vskip 10pt
In this paper, we have shown that deterministic chaos
that emerges in dynamical systems as a result of iterative applications of holomorphic maps can be implemented in the context of Floquet CFT driving.
Making use of the fact that Floquet evolution of a 2d CFT can be expressed as iterated composition of conformal maps followed by analytical continuation to real time, we have studied consequences of choosing the maps that generate dynamical systems exhibiting the order-chaos transition. In the language of Hamiltonian dynamics, it means that the system Hamiltonian is swung between the deformed and undeformed versions at equal periods of time, so that the CFT is perturbed in stroboscopic manner. When the corresponding dynamical system is tuned towards the transition point, self-similar corrections to the entanglement entropy arise on top of the main smooth component, and the  entropy as a function of bipartition coordinate acquires fractal shape. Usually, the entropy scales monotonously with the size of the smaller subsystem, and the scaling law serves as a signature of the class the quantum state belongs to.
In this regard, the situation when entanglement entropy strongly oscillates as a function of the number of degrees of freedom included in the subsystem is highly anomalous. It is also interesting to note that, in our setting, the fractal structures emerge as a result of periodic in time regular driving\footnote{Recent studies discussed similar phenomena in systems driven randomly or quasi-periodically \cite{Wen:2020wee, Lapierre}.}.

Here we analyzed the entropy scaling (see Supplemental Material \cite{SM} for the analysis of one-point correlation functions), and such an unusual structure of quantum correlations would definitely affect transport and spectral properties of the system. Thus it could be important to study states of this kind in more detail because the possibility to create them by means of stroboscopic driving might have unexpected implications for many-body quantum physics.

Since chaotic dynamical systems are ubiquitous, we can expect that quantum states with non-regular behaviour of observables and quantum correlation measures can in fact be engineered in a broad class of Floquet systems. For example, in certain cases $(1+1)$-d conformal field theories can be regarded as continuous limits of spin chains, which, in turn, can be implemented experimentally using cold atoms in optical lattices \cite{Cirac} or by STM manipulation \cite{STM}. Another potentially promising research avenue is to understand whether fractal-generating driving can be realized as some black hole perturbations in the holographic context (see for example \cite{Andrade:2019rpn}), and study its meaning in classical and quantum gravity.

\acknowledgments
We would like to thank P.~Caputa for the valuable comments on the first version of the manuscript and M.~I.~Katsnelson, B.~Lapierre, and X. Wen for useful discussions. The work of D.S.A. which consisted in designing the project and computing energy and entanglement evolution for the tent map was supported by the Russian Science Foundation through grant 17-71-20154  and performed at the Steklov Mathematical Institute of Russian Academy of Sciences. The work of A.A.B., who computed Hausdorff dimensions of the entropy curves and wrote the manuscript, was supported by the Russian Science Foundation through grant 18-12-00185. A.A.I. acknowledges financial support from Dutch Science Foundation NWO/FOM under Grant No. 16PR1024.  The work was partailly supported by Knut and Alice Wallenberg Foundation through Grant No. 2018.0060.

\appendix


\begin{thebibliography}{100}
\bibitem{Jiang:2011xv}
L.~Jiang, T.~Kitagawa, J.~Alicea, A.~Akhmerov, D.~Pekker, G.~Refael, J.~I.~Cirac, E.~Demler, M.~D.~Lukin and P.~Zoller,
``Majorana Fermions in Equilibrium and Driven Cold Atom Quantum Wires,''
Phys.\ Rev.\ Lett.\  \textbf{106}, 220402 (2011)

\bibitem{Wieder:2020lvf}
B.~J.~Wieder, Z.~Wang, J.~Cano, X.~Dai, L.~M.~Schoop, B.~Bradlyn and B.~A.~Bernevig,
``Strong and fragile topological Dirac semimetals with higher-order Fermi arcs,''
Nature Commun.\  \textbf{11}, no.1, 627 (2020)

\bibitem{ab1}
P.~Ponte, Z.~Papi\'c, F.~Huveneers, D.~A.~Abanin, ``Many-Body Localization in Periodically Driven Systems,'' Phys.\ Rev.\ Lett.\ \textbf{114}, 140401 (2015)

\bibitem{ab2}
D.~A.~Abanin, W.~ De Roeck, and F.~ Huveneers, ``Theory of Many-Body Localization in Periodically Driven Systems,'' Annals of Physics 372 (2016)

\bibitem{Maffei:2018knw}
M.~Maffei, A.~Dauphin, F.~Cardano, M.~Lewenstein and P.~Massignan,
``Topological characterization of chiral models through their long time dynamics,''
New J.\ Phys.\  \textbf{20}, no.1, 013023 (2018)

\bibitem{rig1}
 L.~ D'Alessio, M.~ Rigol, ``Long-Time Behavior of Isolated Periodically Driven Interacting Lattice Systems,'' Phys.\ Rev.\ X \textbf{4}, 041048 (2014)

\bibitem{Stepanov}
E.~A.~Stepanov, C.~Dutreix, and M.~I.~Katsnelson, ``Dynamical and Reversible Control of Topological Spin Textures'',
Phys. Rev. Lett. {\bf 118} (2017) 157201

\bibitem{Po:2016qlt}
H.~C.~Po, L.~Fidkowski, T.~Morimoto, A.~C.~Potter and A.~Vishwanath,
``Chiral Floquet Phases of Many-Body Localized Bosons,''
Phys.\ Rev.\ X \textbf{6}, no.4, 041070 (2016)

\bibitem{Wen:2018agb}
X.~Wen and J.~Wu,
``Floquet conformal field theory,''
[arXiv:1805.00031 [cond-mat.str-el]].

\bibitem{Fan:2019upv}
R.~Fan, Y.~Gu, A.~Vishwanath and X.~Wen,
``Emergent Spatial Structure and Entanglement Localization in Floquet Conformal Field Theory,''
[arXiv:1908.05289 [cond-mat.str-el]].

\bibitem{Lapierre:2019rwj}
B.~Lapierre, K.~Choo, C.~Tauber, A.~Tiwari, T.~Neupert and R.~Chitra,
``Emergent Black Hole Dynamics in Critical Floquet Systems,''
Phys.\ Rev.\ Research \textbf{2}, 023085 (2020)

\bibitem{Hikihara:2011}
T. Hikihara and T. Nishino, ``Connecting distant ends of one-dimensional critical systems by a sine-square deformation,'' Phys.\ Rev.\ B \textbf{83}, 060414(R) (2011)

\bibitem{MacCormack:2018rwq}
I.~MacCormack, A.~Liu, M.~Nozaki and S.~Ryu,
``Holographic Duals of Inhomogeneous Systems: The Rainbow Chain and the Sine-Square Deformation Model,''
J. Phys. A \textbf{52}, no.50, 505401 (2019)

\bibitem{Wen:2016inm}
X.~Wen, S.~Ryu and A.~W.~W.~Ludwig,
``Evolution operators in conformal field theories and conformal mappings: Entanglement Hamiltonian, the sine-square deformation, and others,''
Phys.\ Rev.\ B \textbf{93}, no.23, 235119 (2016)


\bibitem{Langmann:2019nna}
E.~Langmann and P.~Moosavi,
``Diffusive Heat Waves in Random Conformal Field Theory,''
Phys.\ Rev.\ Lett. \textbf{122}, no.2, 020201 (2019)

\bibitem{Caputa:2020mgb}
P.~Caputa and I.~MacCormack,
``Geometry and Complexity of Path Integrals in Inhomogeneous CFTs,''
[arXiv:2004.04698 [hep-th]].

\bibitem{Moosavi:2019fas}
P.~Moosavi,
``Inhomogeneous conformal field theory out of equilibrium,''
[arXiv:1912.04821 [math-ph]].


\bibitem{Gawedzki:2017woc}
K.~Gawedzki, E.~Langmann and P.~Moosavi,
``Finite-time universality in nonequilibrium CFT,''
J. Statist. Phys. \textbf{172}, 353 (2018)

\bibitem{coh}
A. ~Lazarides, A.~Das, and R.~Moessner, ``Periodic thermodynamics of isolated quantum systems,'' Phys.\ Rev.\ Lett. \textbf{112}, 150401 (2014)

\bibitem{Brown:DynSys}
R.~J.~Brown, ``A Modern Introduction to Dynamical Systems,'' Oxford University Press, 2018

\bibitem{dyn1}
I.~Cornfeld, S.~Fomin, and Y.~Sinai, ``Ergodic theory,'' Vol. 245. Springer Science \& Business Media, 2012.

\bibitem{dyn2}
A.~Katok, and B.~Hasselblatt, ``Introduction to the modern theory of dynamical systems,'' Vol. 54. Cambridge university press, 1997.

\bibitem{Hofstadter:1976}
D.~ R.~ Hofstadter, ``Energy levels and wavefunctions of Bloch electrons in rational and irrational magnetic fields,'' Phys.\ Rev.\ B \textbf{14}, no.6, 2239-2249 (1976)

\bibitem{Kohmoto:1983}
M.~Kohmoto, L.~P.~Kadanoff, and C.~Tang, ``Localization Problem in One Dimension: Mapping and Escape,'' Phys. Rev. Lett. \textbf{50}, 1870 (1983)

 \bibitem{Domany_1983}
E.~{Domany}, S.~{Alexander}, D.~{Bensimon}, and L.~P.~{Kadanoff}, ``Solutions to the Schr\"odinger equation on some fractal lattices,'' Phys.\ Rev.\ B \textbf{28}, 3110 (1983)

\bibitem{ratiter}
A. F. Beardon, ``Iteration of rational functions,'' Springer-Verlag, 1991.

\bibitem{Shang_etal:2015}
	J.~Shang, Y.~Wang, M.~Chen, J.~Dai, X.~Zhou, J.~Kuttner, G.~Hilt, X.~Shao, .~M. Gottfried, and K.~Wu, ``Assembling molecular Sierpinski triangle fractals,'' Nature Chemistry \textbf{7}, 389 (2015)
	
\bibitem{Kempkes_etal:2018}
	S. N. Kempkes, M. R. Slot, S. E. Freeney, S. J. M. Zevenhuizen, D. Vanmaekelbergh, I. Swart, and C. Morais Smith, ``Design and characterization of electrons in a fractal geometry,'' Nature Physics \textbf{15}, 127 (2018)

\bibitem{Gubser:2018}
S. S. Gubser, C. Jepsen, Z. Ji, and B. Trundy, ``Continuum limits of sparse coupling patterns,''
Phys.\ Rev.\ D \textbf{98}, 045009 (2018)

	\bibitem{Westerhout:2018}
	T.~Westerhout, E.~van Veen, M.~I. Katsnelson, and S.~Yuan, ``Plasmon confinement in fractal quantum systems,'' Phys.\ Rev.\ B \textbf{97}, 205434 (2018).

	\bibitem{Pal_etal:2019}	
	S. Manna, B. Pal, W. Wang, A. E. B. Nielsen, ``Anyons and Fractional Quantum Hall Effect in Fractal Dimensions,'' Phys. Rev. Research \textbf{2}, 023401 (2020)

\bibitem{IlKatSheng:2019}
	A.~A.~Iliasov, M.~I.~Katsnelson, and  S.~Yuan,
	``Power-law energy level spacing distributions in fractals,'' Phys.\ Rev.\ B, \textbf{99}, 075402 (2019)

\bibitem{IlKatSheng:2020}
	A.~A.~Iliasov, M.~I.~Katsnelson, and  S.~Yuan, ``Hall conductivity of a Sierpinski carpet,'' Phys. Rev. B \textbf{101}, 045413 (2020)
	
\bibitem{IlKatSheng:2020b}	
	A.~A.~Iliasov, M.~I.~Katsnelson, and S.~Yuan, ``Linearized spectral decimation in fractals,'' arXiv:2006.02339 [cond-mat.dis-nn].

\bibitem{chaosrmt1}
 T.~H.~Seligman, J.~J.~M.~Verbaarschot, M.~R.~Zirnbauer,  ``Quantum spectra and transition from regular to chaotic classical motion,'' Phys.\ Rev.\ Lett. \textbf{53} no.3, 215 (1984)

\bibitem{chaosrmt2}
 T.~H.~Seligman, J.~J.~M.~Verbaarschot, M.~R.~Zirnbauer, ``Spectral fluctuation properties of Hamiltonian systems: the transition region between order and chaos,'' J. Phys. A: Math. Gen. \textbf{18} no.14, 2751 (1985)

\bibitem{dyn3}
S.~ H. Strogatz, ``Nonlinear dynamics and Chaos,''
Addison-Wesley Pub, 1994.

\bibitem{Wen:2018vux}
X.~Wen and J.~Q.~Wu,
``Quantum dynamics in sine-square deformed conformal field theory: Quench from uniform to nonuniform conformal field theory,''
Phys.\ Rev.\ B \textbf{97}, no.18, 184309 (2018)
\bibitem{SM} Supplemental Material of this paper.

\bibitem{Curtright:2009}
  T.~Curtright and C.~Zachos, ``Evolution profiles and functional equations,'' J. Phys. A: Math. Theor. \textbf{42}, 485208 (2009)

\bibitem{Milnor:2004}
    J. W. Milnor, ``On Latt{\'es} Maps'', arXiv:0402147 [math.DS]

\bibitem{CollJeanPierre:1980}
    P.~Collet, J.-P.~Eckmann, ``Iterated Maps on the Interval as Dynamical Systems,'' (1980)

\bibitem{MilnThurst:1988}
Milnor, John W.; Thurston, William (1988), "On iterated maps of the interval", Dynamical systems (College Park, MD, 1986–87), Lecture Notes in Mathematics, 1342, Berlin: Springer, pp. 465–563, MR 0970571

\bibitem{Saleur} R.~Couvreur, J.~L.~Jacobsen, and H.~Saleur, ``Entanglement in nonunitary quantum critical spin chains,'' Phys. Rev. Lett. {\bf 119} (2017): 040601

\bibitem{Moosavi}P.~Moosavi, ``Inhomogeneous conformal field theory out of equilibrium,'' arXiv:1912.04821

\bibitem{WuAn}H.~Wu and J.~H.~An, ``Floquet topological phases of non-Hermitian systems,'' Phys. Rev. B {\bf 102} (2020) 041119(R)

\bibitem{GongWang}J.~Gong and Q.~H.~Wang, ``Stabilizing non-Hermitian systems by periodic driving,''
Phys. Rev. A {\bf 91} (2015) 042135

\bibitem{Calabrese:2004eu}
P.~Calabrese and J.~L.~Cardy,
``Entanglement entropy and quantum field theory,''
J. Stat. Mech. \textbf{0406}, P06002 (2004)

\bibitem{Calabrese:2009qy}
P.~Calabrese and J.~Cardy,
``Entanglement entropy and conformal field theory,''
J. Phys. A \textbf{42}, 504005 (2009)

\bibitem{HeadrickTakayanagi} M. Headrick and T. Takayanagi, ``Holographic proof of the strong subadditivity of entanglement entropy,'' Phys. Rev. D \textbf{76} (2007) 106013

\bibitem{Dubuc}B.~Dubuc, J. F.~Quiniou, C.~Roques-Carmes, C.~Tricot, and S. W.~Zucker,  ``Evaluating the fractal dimension of profiles,'' Phys. Rev. A, 39(3) (1989) 1500

\bibitem{Pilgrim}I.~Pilgrim and R.~P.~Taylor, ``Fractal analysis of time-series data sets: Methods and challenges,'' Fractal Analysis, IntechOpen, 2018

\bibitem{Wen:2020wee}
X.~Wen, R.~Fan, A.~Vishwanath and Y.~Gu,
``Periodically, Quasi-periodically, and Randomly Driven Conformal Field Theories: Part I,''
[arXiv:2006.10072 [cond-mat.stat-mech]].

\bibitem{Lapierre}
B.~ Lapierre, K.~ Choo, A.~ Tiwari, C.~ Tauber, T.~ Neupert, R.~ Chitra,
``The fine structure of heating in a quasiperiodically driven critical quantum system,''
[arXiv:2006.10054 [cond-mat.str-el]].

\bibitem{Cirac}
J.~J.~Garcia-Ripoll, M.~A.~Martin-Delgado, and J.~I.~Cirac, ``Implementation of Spin Hamiltonians in Optical Lattices,''
Phys.\ Rev.\ Lett. \textbf{93}, 250405 (2004)

\bibitem{STM}
C.~F.~Hirjibehedin, C.~P.~Lutz, A.~J.~Heinrich,
``Spin Coupling in Engineered Atomic Structures,'' Science \textbf{312}, 5776, pp. 1021-1024 (2006)

\bibitem{Andrade:2019rpn}
T.~Andrade, C.~Pantelidou, J.~Sonner and B.~Withers,
``Driven black holes: from Kolmogorov scaling to turbulent wakes,''
[arXiv:1912.00032 [hep-th]].

\end{thebibliography}
\end{document}


\title{Supplemental material for ``Deterministic chaos and fractal entropy scaling in 2d Floquet CFT''}
\author{Dmitry S. Ageev}
\author{Andrey A. Bagrov}
\author{Askar A. Iliasov}

\maketitle
\section{Sine-square deformation}
While in this paper, we focus on the CFT deformations corresponding to the driving induced by the logistic and tent maps, for completeness we would like to briefly discuss the so-called regularized sine-square deformation (rSSD) which was studied in the previous works on Floquet driving \cite{Wen:2018agb}. This deformation is given by the following choice of $f(w)$ in Eq. (1) of the main text:
\be 
f(w)=f_{rSSD}(w)=1-\cosh\frac{2\pi w}{L}\tanh(2\theta).
\ee 
 Here, $\theta$ is the regularization parameter, and $\theta \rightarrow \infty$ gives the sine-square deformation (SSD) with $f_{SSD}(w)=-2\sinh^2(\pi w/L)$. This case is exactly solvable, and the corresponding one-step driving function $z_1$ is just the M\"obius transformation
 \be 
z_1^{\text{Mob}}=\frac{a z+b}{c z +d},
 \ee 
 where $a,b,c$, and $d$ depend on $L$, $\theta$ and the duration of one step $\tau$.
 
The composition of $n$ M\"obius transformations is again a M\"obius transformation, and the analysis of such dynamics reduces to the analysis of fixed points of $z_n$ defined by parameters $a,b,c,d$ \cite{mob}. Depending on the choice of parameters, the driven CFT enters either heating or non-heating phase. In the heating phase, the existence of two finite fixed points of the M\"obius transformation induces energy density accumulation in two points in the coordinate space of the CFT, and the system heats up in these point unlimitedly over the course of driving. 
Here, we study deformations corresponding to classical dynamical systems exhibiting deterministic chaos that are very different from (r)SSD. In our case, the number of energy concentration points can be made arbitrary large by taking the driving dynamical system close to the chaotic regime. This is similar to what can be observed in randomly or quasi-periodically driven systems \cite{Wen:2020wee, Lapierre}. However, what we observe should not be called a heating phase, since the energy density in concentration points remains finite even when the system is driven for arbitrary long time (at least for the tent type of driving). 
\section{Elements of the theory of dynamical systems}
Here we provide some background information about dynamical systems defined by compositions of holomorphic functions which are an important ingredient of our considerations.

In the theory of dynamical systems, the commonplace is to trivialize the iterative evolution by a coordinate change. 
If the evolution is governed by some fixed map $h(x)$, one can perform coordinate transformation obeying 
\be
\Psi(h(x))=\lambda^t \Psi(x), \label{eq:Shroeder}
\ee
where $t$ is a continuous parameter, so that the resulting evolution of the dynamical system reduces to a mere dilatation. When $t$ is integer it corresponds to the number of iterations of the dynamical system. Eq. \eqref{eq:Shroeder} is known in the theory of dynamical systems as the Schr{\"o}der equation.
 
It is easy to recognise the Schr{\"o}der equation in Eq. (5) of the main text. To prove the existence of solutions to the Schr{\"o}der equation is a difficult problem for a general map \footnote{It is worth to notice that the Schr{\"o}der equation is a topological semi-conjugate relation between dynamical systems. Semi-conjugation does not establish an equivalence relation between dynamical systems (which is given by topological conjugation). Nevertheless, it still preserves some topological properties e.g. topological transitivity that can be viewed as a measure of complexity of the dynamics \cite{Brown:DynSys}.}. Nevertheless, it is known that for certain univalent maps (with one maximum in the unit interval), such as the tent and the logistic maps, the Schr{\"o}der equation admits explicit solutions \cite{Curtright:2009, Milnor:2004}.

In the context of Floquet CFT, the tent and the logistic maps are interesting because they are the simplest and canonical examples of dynamical systems showing the transition to the regime known as deterministic chaos. This regime demonstrates highly non-regular behaviour that is apparently random and amenable for studying with probabilistic techniques, and nevertheless can be described by rather simple deterministic rules. 

Let us briefly remind the mechanism of the deterministic chaos transition. A point satisfying
\be
x_0=h_{m}(x_0)\equiv\underbrace{(h\circ  h \ldots \circ h)}_{m\,\, times}(x_0),
\ee
identity is called a periodic point of the dynamical system with period $m$. The order-chaos transition can then be understood in terms of period-doubling bifurcations, in which a small deformation of $h$ map leads to a new behavior with twice the period of the original system. The system becomes effectively chaotic when its period is large enough.

For the logistic map, the structure of minima and maxima of $h_{m}$ changes during the iterative process. In the trivial regime of small $\alpha$, all the iterated maps look qualitatively the same and have just one maximum. In the truly chaotic regime ($\alpha=4$), every iteration doubles the number of maxima, and, in the limit of infinite number of iterations, the graph of the limiting function $h_{m\to\infty}$ has Hausdorff dimension 2 and covers the unit square. In the intermediate regime, the Hausdorff dimension lies somewhere between 1 and 2, and graph of the limiting function forms a fractal.
$\,$

The tent map is piecewise-linear and thus is easier to analyze. For example, one can see that in the regime of deterministic chaos, $\beta=2$, the $n$-th iteration of the tent map is just $2^n$ shrunk copies of the original map:
\begin{gather} \label{tentiter}
z_n^{\text{tent}}(x)=\begin{cases}
2^{n} (x-\frac{k}{2^{n-1}}), \, x\in(\frac{k}{2^{n-1}},\frac{2k+1}{2^{n}});\\
2^{n} (\frac{k+1}{2^{n-1}}-x), \, x\in(\frac{2k+1}{2^{n}},\frac{k+1}{2^{n-1}});
\end{cases}\\
 k=0,1,2,\ldots(2^{n-1}-2),(2^{n-1}-1) \nonumber
\end{gather}
It is interesting to note that this case is topologically conjugated to the chaotic phase of dynamical system defined by the logistic map with doubling maxima \cite{CollJeanPierre:1980}.

\begin{figure}[h]
        \centering
         \includegraphics[width=7cm]{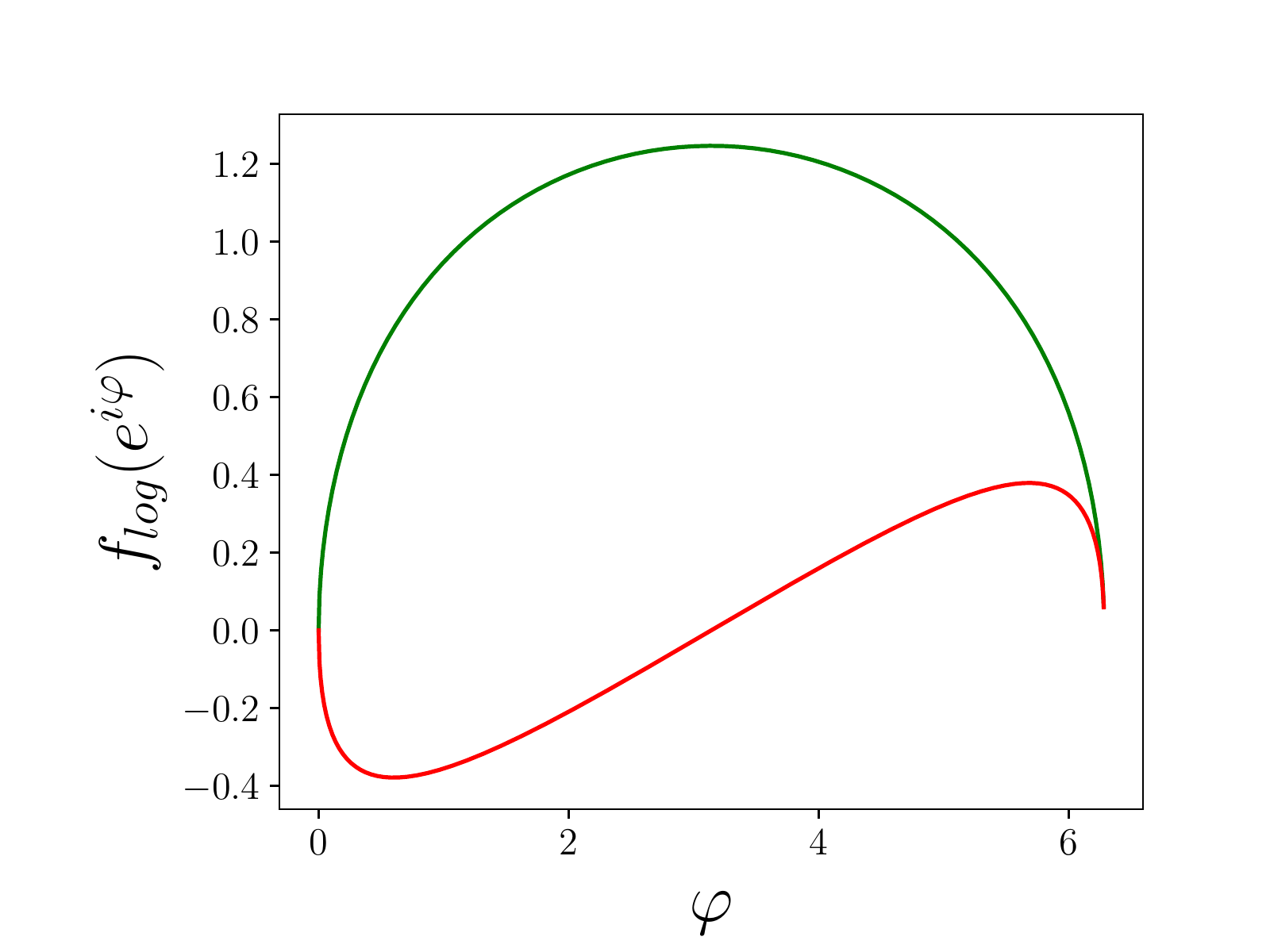}
        \caption{\label{fig:f_profile} Real (green) and imaginary (red) parts of deformation function $f_{log}(z)$ corresponding to the logistic type of driving at $\alpha=\alpha_{chaotic}=4$, evaluated on the unit circle.}        
    \end{figure}
As discussed in the main text, dynamical systems of this kind correspond to the evolution of a conformal field theory after a quench by some deformed Hamiltonian, Eqs. (1) - (6) of the main text. In order to reformulate the Floquet CFT in terms of iterated holomorphic maps, we need to consider a composition of two maps (Eq. (7) of the main text): dilatation $z_1$ (which corresponds to undeformed CFT), and map defining the dynamical system $\widetilde z_1$. It is more convenient to start evolution with deformed CFT step $\widetilde z_1$.
It is easy to show that composition $z_1 \circ \tilde{z}_1$ is equal to a single logistic or tent map, but with rescaled $\alpha$ or $\beta$ 
\begin{eqnarray}
z_1^{\text{logistic}}\rightarrow e^{\frac{2\pi \tau}{L}}\cdot z^{\text{logistic}}_1,\,\,\,\,\,\,\alpha \rightarrow  e^{\frac{2\pi \tau}{L}} \alpha,\\
z_1^{\text{tent}}\rightarrow e^{\frac{2\pi \tau}{L}}\cdot z^{tent}_1,\,\,\,\,\,\,\beta \rightarrow  e^{\frac{2\pi \tau}{L}} \beta,
\end{eqnarray}
where $\tau$ is the time step of the undeformed Hamiltonian. Hence the Floquet evolution given by Eq. (7) of the main text is identical to the CFT evolution following a single Hamiltonian quench with deformation corresponding to the rescaled map.
\section{Deformation functions at the transition point}
While we do not use deformation functions $f(z)$ explicitly in our calculations, it is still instructive to visualize them. For the logistic map at $\alpha=\alpha_{chaotic}$, in z coordinates it is
\begin{eqnarray}
f_{log}(z) = \frac{\chi(z)}{z \chi'(z)}=\sqrt{\frac{1}{z}-1}\arcsin(\sqrt{z}),
\end{eqnarray}
which is plotted in Fig.\ref{fig:f_profile}. In the domain of integration ($(0, 1)$ interval in $w$ coordinates or the unit circle in $z$ coordinates), it is a complex-valued periodic function.

For the tent map, the situation is a bit trickier. Along the unit circle, it is just a constant:
\begin{eqnarray}
f_{tent}(e^{i\varphi})=0.5, \,\,\,\varphi\in (0, \pi)\cup(\pi, 2\pi),
\end{eqnarray}
with a removable singularity at $\varphi=\pi$,
but due to branching points of $\arcsin$ function at the end of the interval, non-trivial dynamics is still induced. It could be desirable to have a Hamiltonian described by smooth deformation function $f(z)$ with properties similar to that of the tent case. For that, one needs to regularize tent map by smoothing out the singular points at the top and at the base of the $\arcsin(\sin(\pi x))$ ``sawtooth'' e.g., by imposing $z^{\text{tent}}_{\text{reg}}(x)= \frac{\pi\beta}{4}(1-\sin(2\pi x+\pi/2))$.
\section{Deformation functions away from the transition point}
\subsection{Logistic map}
In the case of logistic map, solution to the Schr{\"o}der equation can be written as a formal series in the neighborhood of $z=0$ \cite{Curtright:2010}:
\be
\chi^{-1}(z,\alpha)=z+z\sum^{\infty}_{n=1}z^n c_{n}(\alpha) \label{eq:rec_series}
\ee
where $c_n$ coefficients are given by the recursion relation:
\be
c_{n+1}=\frac{1}{1-\alpha^{n+1}}\sum^{n}_{j=0}c_{n-j}c_{j},
\ee
and the corresponding eigenvalue of the solution is equal to parameter $\alpha$ of the map. Convergence of the series depends on  $\alpha$. In \cite{Curtright:2010}, this series has been extensively studied, and it was suggested that $\chi^{-1}$ is an entire function for $\alpha \in [2, 4]$.
	
The general behaviour of this series is quite complicated, however, in certain regimes it admits closed solutions (see \cite{Curtright:2010} for the details of derivation). For $\alpha=2$:
\be
\chi(z,2)=-\frac{1}{2}\ln(1-2z),
\ee
and corresponding deformation of the Hamiltonian is
\be
f(z)=(1-\frac{1}{2z})\ln(1-2z),	\ee
which has a branching point within a unit circle.
	
In the limit of small $\alpha$, where there is no chaotic behaviour of entanglement entropy, by taking into account only linear term in $\alpha$ we obtain
\be
\chi(z,\alpha)=z(1-z)(1-\alpha z(1-z))
\ee
with the corresponding deformation
\be
f(z)=\frac{(1-z)(1-\alpha z(1-z))}{(1-2z)(1-2\alpha z(1-z))}
\ee
While there is no proof at this moment, one could possibly conjecture that, in the case of logistic map, emergence of the fractal structures is connected with existence of branching points in the Hamiltonian deformation function.
    
\subsection{Tent map}
For the tent map driving, the corresponding deformations of the Hamiltonian can be derived from their logistic cousins by means of semiconjugate relation via Milnor-Thurston kneading theory  \cite{MilnThurst:1988}. The full analysis is rather cumbersome, and, since we in fact do not need the explicit expressions for $f(z)$, here we just ouline the pathway towards constructing them.

To do that, one first needs to define the so-called kneading coordinates $\theta_n$ on $[0,1]/\Gamma$, where $\Gamma$ is a set of preimages of $\frac{1}{2}$. Starting with
\be
\theta_0 (x)=\begin{cases}+1, \text{if  } x<\frac{1}{2} \\ -1, \text{if  } x>\frac{1}{2},  \end{cases} 
\ee
the next coordinates are defined by the recursive relation $\theta_{n}(x)=\theta_{n-1}(x)\theta_0(f^n(x))$, where $f$ is the logistic map.

Using $\theta_n(x)$, one constructs a formal series:
\be
\theta(x,t)=\sum^{\infty}_{n=0}\theta_n (x)t^n.
\ee
At the next step, function $\theta(x^{-})$ on the whole unit interval $(0,1)$ is defined as the limit through elements $y\in[0,1]/\Gamma$:
\be
\theta(x^{-},t)=\lim_{y\nearrow x}\theta(y,t).
\ee
The semi-conjugate relation of tent map with parameter $\beta$ to logistic map within the unit interval $(0,1)$ is then given by the following expression:
\begin{gather}
g(x)=\frac{1}{2}\left(1-\left(1-\frac{1}{\beta}\right)\theta\left(x^{-},\frac{1}{\beta}\right)\right),\\
z_{log} = g^{-1} \circ z_{tent} \circ g.    \nonumber
\end{gather}
The corresponding deformation of Hamiltonian can then be reconstructed from series \eqref{eq:rec_series}.
\section{On Hermicity of the deformations}
We are dealing with Hamiltonian on a strip of width $L$:
\be \label{H0}
H=\int_0^L dx\left(f(w)T(w)+f(\bar w) \bar T(\bar w)   \right),
\ee 
where $w=\tau+ ix$, $x\in (0,L)$.
On $(0,L)$ interval, $f(x)$ is a function of real argument that can be 
expanded as a Fourier series:
\be \label{f:Fourier}
f(x)=\sum_n c_n e^{2\pi inx/L},
\ee
and analytically continued to the complex plane:
\be \label{f:Fourier_an}
f(x)=\sum_n c_n e^{2\pi(n\tau+inx)/L}=\sum_n c_n e^{2\pi nw/L}
\ee
Making $z=e^{2\pi w/L}$ transformation, we obtain the Hamiltonian in new coordinates:
\be \label{H0_z}
H=\frac{1}{iL}\oint zdz \sum_n c_n z^nT(z)+ c.c.,
\ee 
or, in terms of the Virasoro generators,
\be \label{H0_L}
H=\frac{1}{L}\sum_n c_n (L_n+\bar L_n).
\ee
For unitary representations of the Virasoro algebra, $L^{\dagger }_n= L_{-n}$, hence Hamiltonian \eqref{H0} is Hermitian if
\be \label{Hermicity_cond}
c_n=c_{-n}.
\ee
For real function $f(x)$, this condition is met trivially and $c_n\in\mathbb{R}$ for all $n$. However, \eqref{Hermicity_cond} can still hold true for a certain choice of complex coefficients $c_n$.
	
Since all the exponents in Eq. \eqref{f:Fourier} are functionally independent, this leads to an explicit condition on $f(x)$. Namely, analytical continuation of the function on $(-L,L)$ interval should possess the following property:
\be
f(x)=f(-x)
\ee
or, in $z$ coordinates:
\be
f(z)=f\big(\frac{1}{z}\big).
\ee
From that one can readily deduce that deformations considered in this paper are non-Hermitian.
\vskip 15pt
\section{Computing fractal dimension of the curves}
As we mentioned in the main part of the paper, to compute fractal dimension of a curve $S(l)$ such that the corresponding coordinate $l$ and value $S$ have different meanings, plain box-counting algorithm is of no use. Instead, we employ the method suggested in \cite{Dubuc}. According to it, to compute fractal dimension of a curve, one first needs to represent it as a discrete series of values $S_i$ of length $N$ (concretely, in our simulations we discretize the curve into $N=5000$ points). Then, for a given $\epsilon$, functions returning the maximal and the minimal values of the series within the $\epsilon$ neighborhood of a given element are defined:
\begin{gather}
    u_\epsilon(i) = \sup\limits_{i'\in R_\epsilon (i)} S_{i'},\\
    b_\epsilon(i) = \inf\limits_{i'\in R_\epsilon (i)} S_{i'}, \nonumber
\end{gather}
where $R_\epsilon (i) = \{i': |i'-i|\leq \epsilon, i' \in [1,N]\}$. $u_\epsilon(i)$ and $b_\epsilon(i)$ are then upper and lower envelops of oscillations of the $S_i$ series at scale set by the given value of $\epsilon$.

The next ingredient is function 
\begin{equation}
V(\epsilon) = \frac{1}{\epsilon^2}\sum\limits_i (u_\epsilon(i)-v_\epsilon(i)),    
\end{equation}
which can be regarded as ``box-counting'' function that returns the number of $\epsilon \times \epsilon$ blocks within the band between two envelops. Fitting it with a power-law
\begin{eqnarray}
V(\epsilon) = \frac{a}{\epsilon^D}, \label{eq:fractal_fit}
\end{eqnarray}
one reconstructs the fractal dimension of the $S$ curve. For the fit, we took a set of values $\epsilon = 2\dots 200$, computed so-called local fractal dimensions for different windows of scales of length $10$ (i.e., by making fit \eqref{eq:fractal_fit} for every range from $\epsilon = 1,\dots 10$ to $\epsilon = 191,\dots 200$), and estimated the fractal dimension to be the mean of local fractal dimensions.

\section{Dependence of fractal dimension on the bipartition coordinate}
To compute fractal dimension of a curve, we have used the whole domain $z\in (0,1)$ as an input for the prescription outlined in the previous section. However, one can naturally ask whether the fractal dimension is stable upon choosing smaller coordinate subintervals. For the values of driving parameter corresponding to the regime where fractal behavior has not fully developed yet (e.g., $\alpha=0.22\alpha_{chaotic}$, middle plot in the bottom panel of Fig. 1 in the main text), it is clearly not the case, as some parts of the entropy profile look like fractals ($l \in (0.25, 0.75)$), while some -- like smooth curves ($l <0.25, \,\,\, l>0.75$). However, once the fractal oscillations cover the whole range of coordinates, there is very little difference between fractal dimension estimates based on different coordinate ranges as long as the range is not too small. For example, for the logistic driving at $\alpha =0.5 \alpha_{chaotic}$, the fractal dimension is $D\simeq 1.87$ when computed on the whole interval, $D\simeq 1.83$ -- on $(0, 0.25)$ and $(0.75, 1)$ intervals, and $D\simeq 1.88$ -- on $(0.25, 0.5)$ and $(0.5, 0.75)$ intervals. For the tent driving at $\beta = 0.65  \beta_{chaotic}$, the fractal dimension estimated on the whole interval is $D\simeq1.526$, estimated on $(0, 0.25)$ and $(0.75, 1)$ intervals -- $D\simeq1.522$, and estimated on $(0.25, 0.5)$ and $(0.5, 0.75)$ intervals -- $D\simeq 1.526$.
\vskip 15pt
\section{Fractal patterns in the energy and momentum density}
\begin{figure*}[t!]
        \centering

        \includegraphics[width=5.7cm]{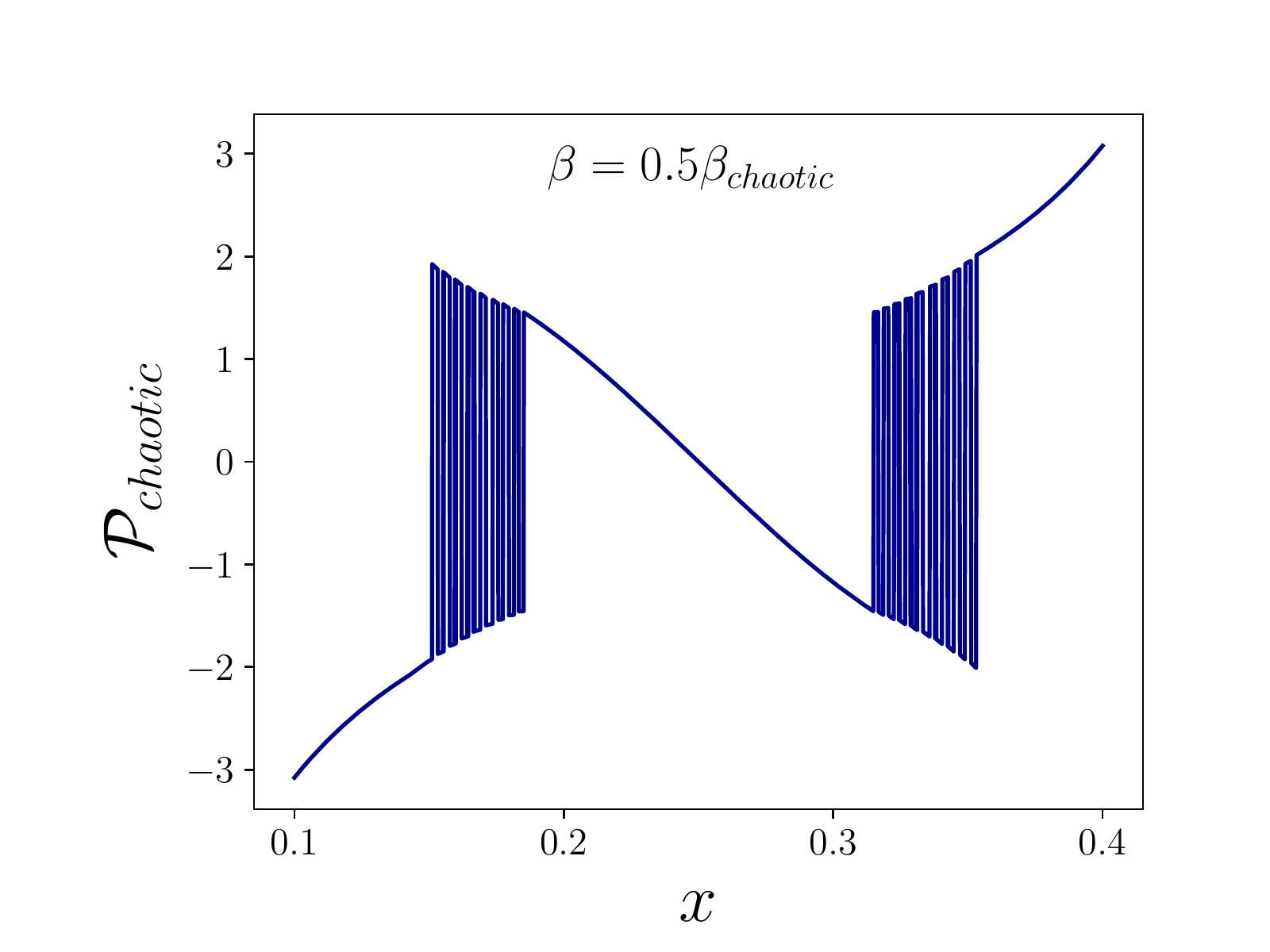}
        \includegraphics[width=5.7cm]{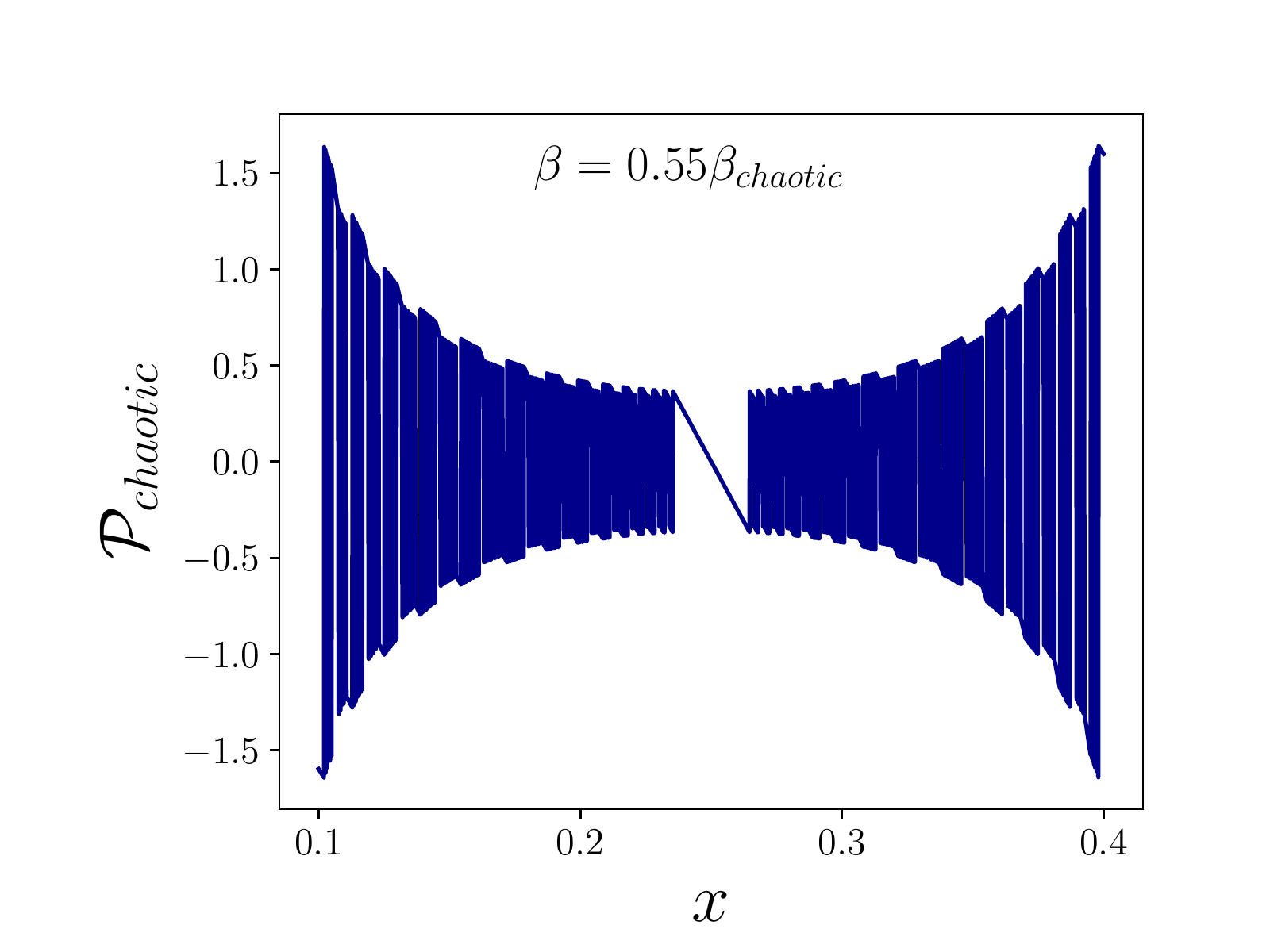}
        \includegraphics[width=5.7cm]{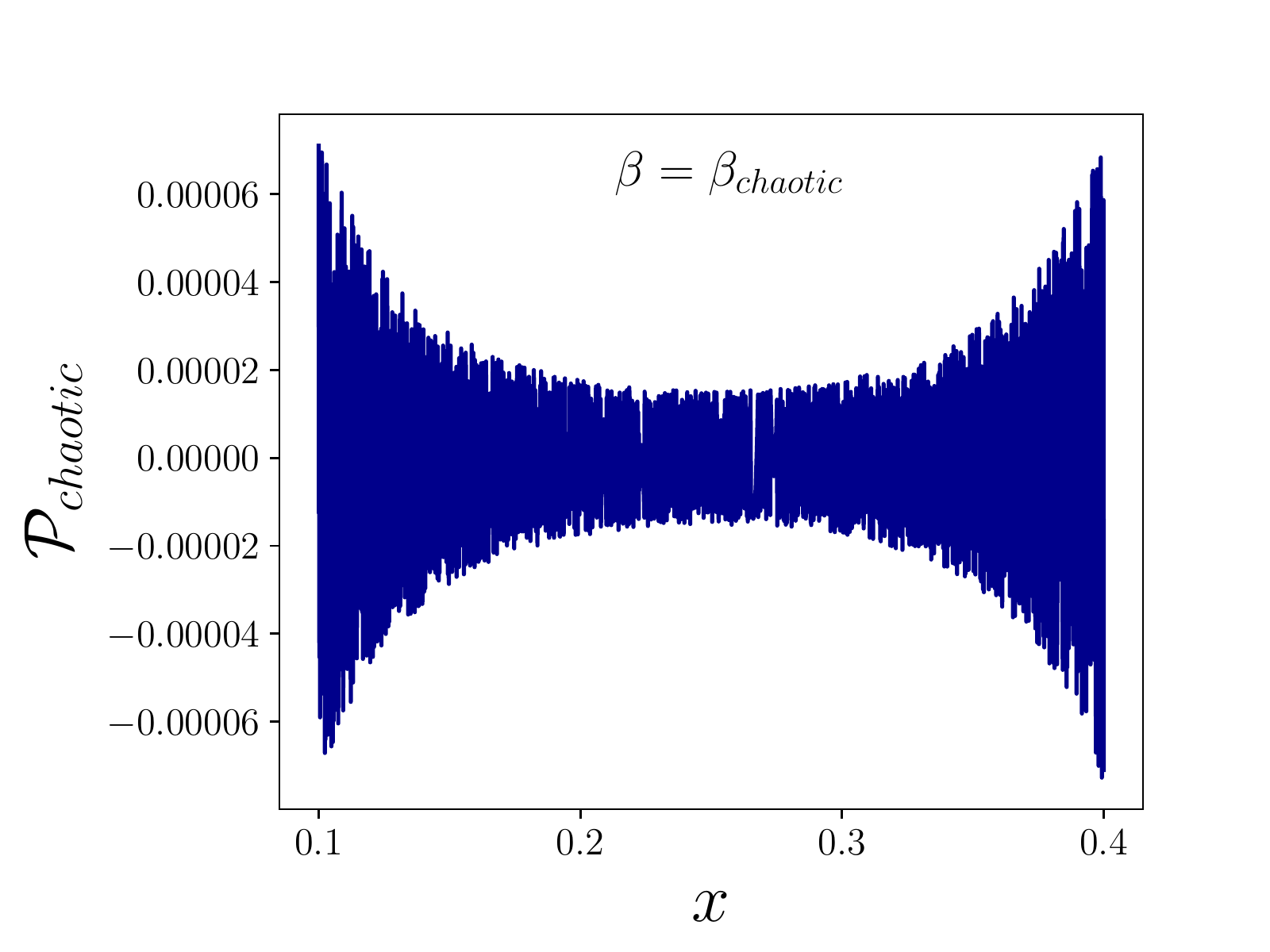}

        \caption{\label{fig:tent-beta} Energy-momentum density ${\cal P}_{chaotic}$  in the interval $x\in(0.1,0.4)$ after 18 steps of driving. Central charge $c=1$ and $L=1$.}        
    \end{figure*}
    \begin{figure*}[t!]
        \centering
         \includegraphics[width=5cm]{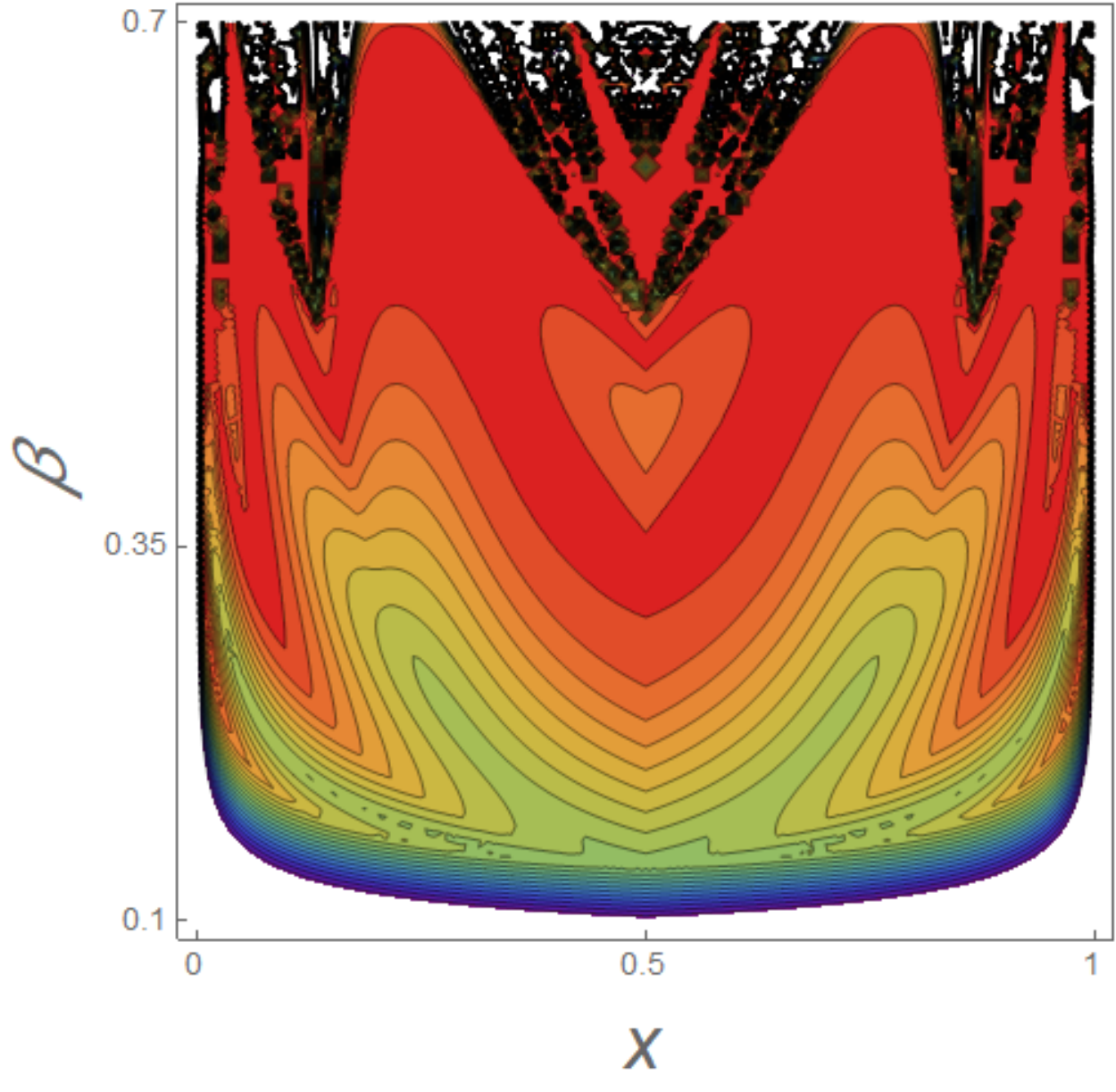}\\
        \includegraphics[width=5.7cm]{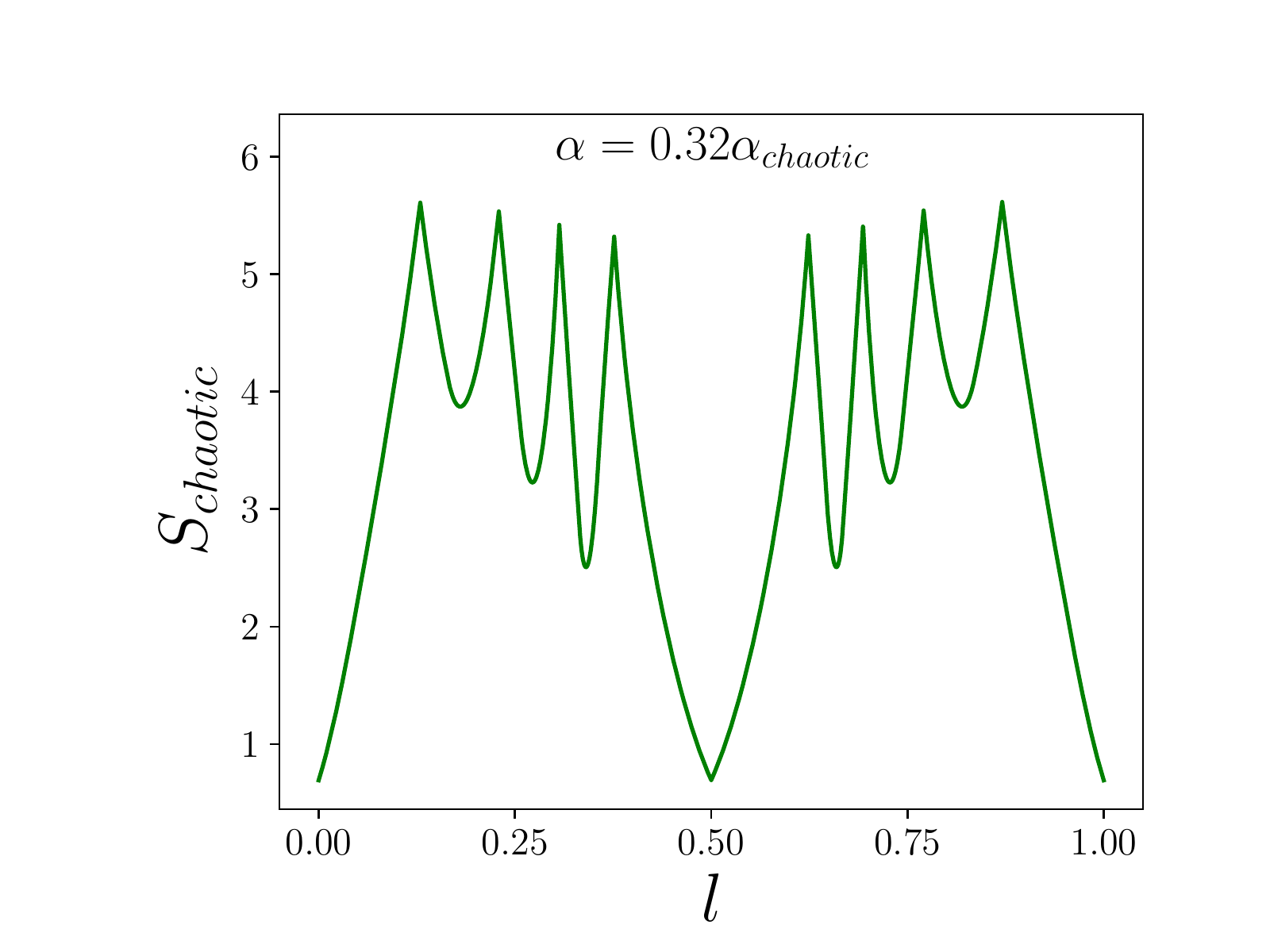}
        \includegraphics[width=5.7cm]{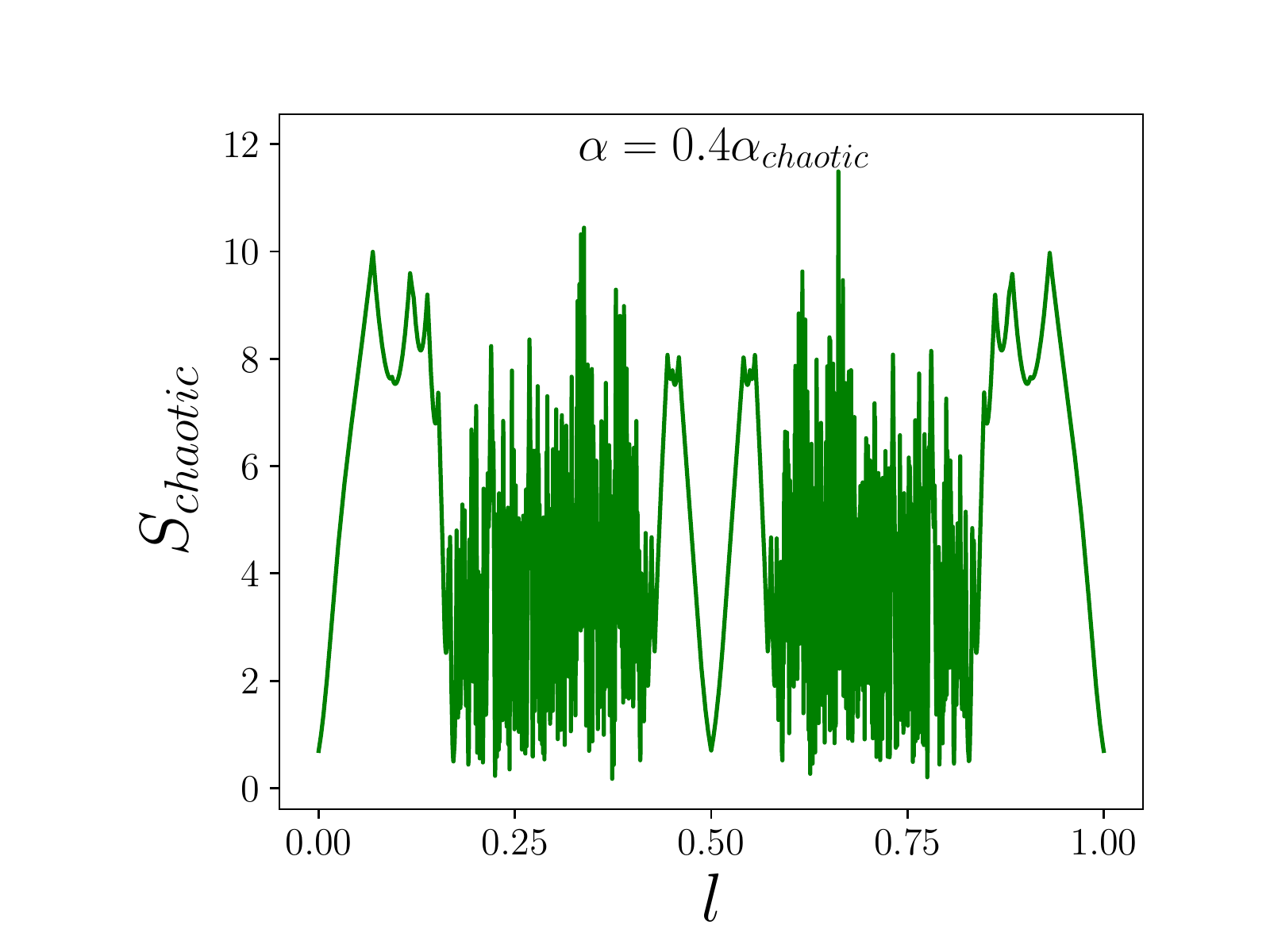}
        \includegraphics[width=5.7cm]{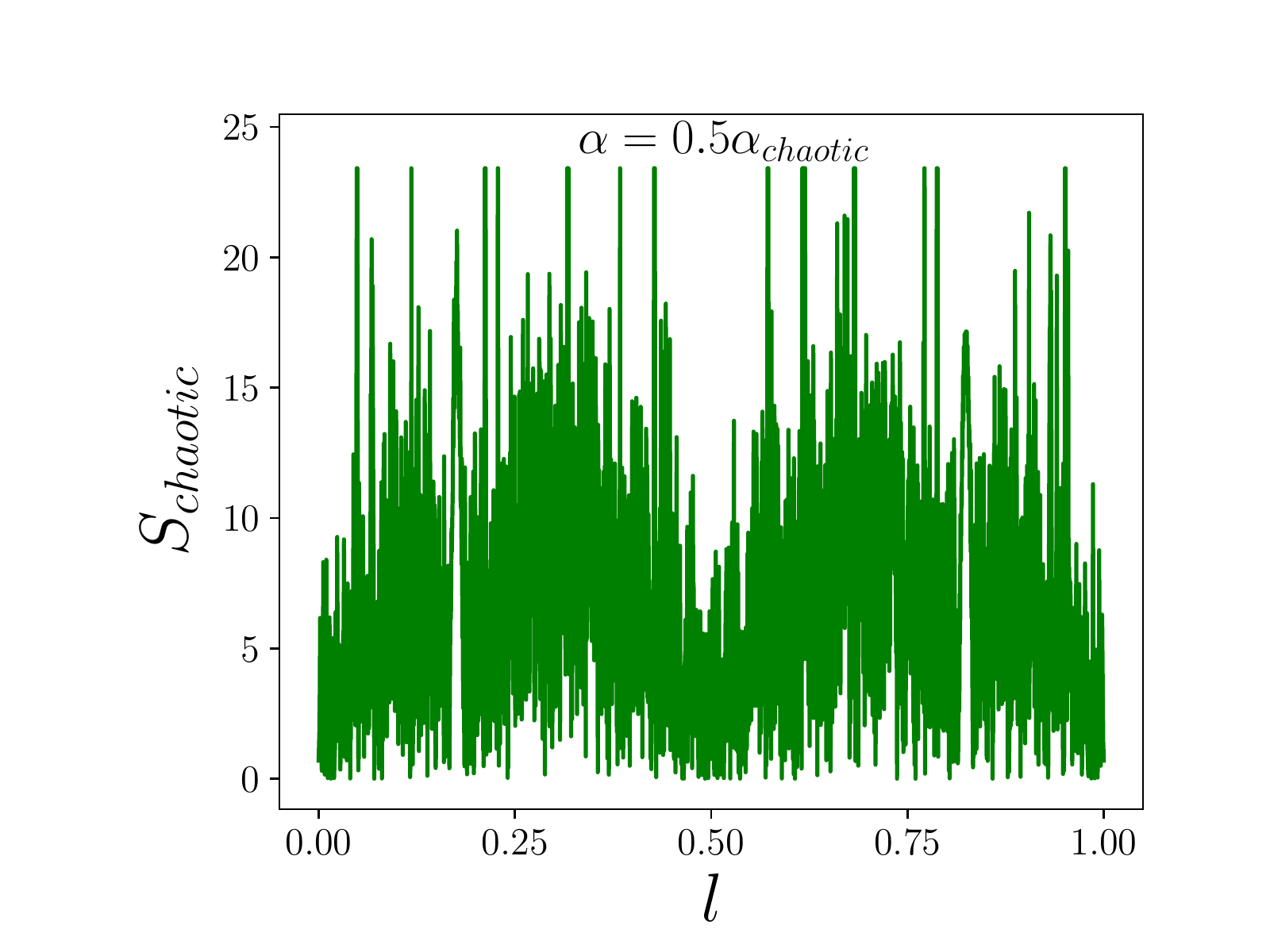}
        \caption{\label{fig:logtentSnorm-alpha} Top: visualization of the combined logistic-tent map profile. Bottom: the irregular part of entanglement entropy $S_{chaotic}$ after $n=7$ combined Floquet steps for the logistic-tent map as a function of bipartition coordinate.}        
    \end{figure*}
In the paper, we put the main focus on studying patterns of quantum correlations encoded in the entanglement entropy. At the same time, it is also instructive to see how the simplest CFT observables behave under the chaos-inducing driving. As an illustrative example, we shall consider evolution of the energy and momentum densities ${\cal E}$ and ${\cal P}$ in the case of tent map. Those are defined as
\begin{eqnarray}
{\cal E}(w,\bar{w})&=&T(w)+\bar T(\bar{w}),\\ {\cal P}(w,\bar{w})&=&i(T(w)-\bar T(\bar{w})). \nonumber 
\end{eqnarray}
Given the piecewise linear nature of the tent map, the same expressions as for the linear map hold true\footnote{
  The deformation corresponding to the tent maps has discontinuities of the derivative on the stripe that could lead to singular terms in the Schwarzian. However, this difficulty can be overcome by using another function with a similar behaviour that has smooth but sharp gradients instead of discontinuities (one of the ways to make a regularization is to consider a polynomial approximation to the tent function).} \cite{Fan:2019upv, Lapierre:2019rwj}: 
\bea\label{eq:tent_energy}
&&{\cal E}(x,t)=\frac{c}{32}\left(\Big(\frac{\partial z_n}{\partial w}\Big)^2\frac{1}{z_n^2}+\Big(\frac{\partial \bar z_n}{\partial w}\Big)^2\frac{1}{\bar z_n^2}\right)-\frac{c\pi^2}{6L^2}\bigg|_{\tau\rightarrow i t}\\
&&{\cal P}(x,t)=i\frac{c}{32}\left(\Big(\frac{\partial z_n}{\partial w}\Big)^2\frac{1}{z_n^2}-\Big(\frac{\partial \bar z_n}{\partial w}\Big)^2\frac{1}{\bar z_n^2}\right)\bigg|_{\tau\rightarrow i t} \nonumber
\eea
where $c$ is the central charge, $z_n$ is the $n$-th step composition of the elementary tent map, and we performed the analytical continuation $\tau\rightarrow i t$.

To make the fractal structures more apparent, it is convenient to rescale the components of stress-energy tensor as
\begin{equation}
T_{\cal N}=\frac{L^2}{4\pi^2}e^{-\frac{4\pi ix}{L}}T,\,\,\,\,\,\,\,\bar T_{\cal N}=\frac{L^2}{4\pi^2}e^{\frac{4\pi ix}{L}}\bar T,
\end{equation}
and redefine
\begin{equation}
{\cal P}_{chaotic}(w)=i(T_{\cal N}-\bar T_{\cal N}).
\end{equation}
In the Lorentzian time, it reads
\begin{gather} 
{\cal P}_{chaotic}(x,t)=\frac{L^2}{4\pi^2}\left(\cos(\frac{4\pi x}{L}){\cal P}(x,t)+\right. \\ \left. \sin(\frac{4\pi x}{L}){\cal E}(x,t)\right) \nonumber
\end{gather} 
This object, while being a bit contrived, serves as a good indicator of fractal dynamics in the CFT.

In Fig. \ref{fig:tent-beta}, we present it after $n=18$ steps of the Floquet driving. It is clear that quite non-trivial irregular oscillating structures emerge already for $\beta=0.5\beta_{chaotic}$ (let us remind that $\beta_{chaotic}=2$). A small increase of $\beta$ (i.e. going up to $\beta/\beta_{chaotic}=0.55$) leads to self-similar duplication of these structures. Further increase of $\beta$ makes the behaviour more and more erratic, and at $\beta=\beta_{chaotic}$ the profile becomes truly chaotic. At the same time, amplitude of ${\cal P}_{chaotic}$ is decreasing upon approaching the chaotic regime.

\section{Combined tent-logistic driving}
For completeness, let us also consider the Floquet CFT dynamics governed by alternating tent/logistic maps:
\be
z_n(z)=\underbrace{(z_{\text{log}}\circ z_{\text{tent}}\circ z_{\text{log}} \ldots \circ z_{\text{tent}})}_{n\,\, times}(z).
\ee
In Fig. \ref{fig:logtentSnorm-alpha}, we present the results for $S_{chaotic}$. We are varying the map parameters $\alpha$ and $\beta$ in such a way that $\alpha/\beta=\alpha_{chaotic}/\beta_{chaotic}$. The behaviour of $S_{chaotic}$ resembles the one of the logistic driving case, also having quite large amplitude, though with a bit different pattern of emergence of self-similar structures.